\documentclass[11pt,a4paper]{article}
\pdfoutput=1
\usepackage{jheppub}
\usepackage{amsmath}
\usepackage{amssymb}
\usepackage{epsfig,multicol,bbm}
\usepackage{graphicx}
\usepackage{bm}
\usepackage{multirow}
\usepackage{bbold}
\usepackage[mathscr]{euscript}
\usepackage{color}
\usepackage{tikz}
\usepackage{book tabs}
\usepackage[latin1]{inputenc}

\usepackage{slashed}

\newcommand{\pf}[2]{\frac{\partial #1}{\partial #2}}

\newcommand{\be}[0]{\begin{equation}}
\newcommand{\ee}[0]{\end{equation}}

\newcommand{\bea}{\begin{eqnarray}}
\newcommand{\eea}{\end{eqnarray}}

\newcommand{\cP}{{\cal P}}

\newcommand{\cD}{{\cal D}}

\newcommand{\nn}{\nonumber}

\newcommand{\abs}[1]{\left\lvert #1\right\rvert}

\DeclareMathOperator{\Tr}{Tr}

\newcommand{\cJ}{\mathcal{J}}

\renewcommand{\log}{\ln}

\newcommand{\veb}[1]{\boldsymbol{\mathrm{#1}}_{\perp}}
\newcommand{\lb}{\Big{\lbrack}}
\newcommand{\rb}{\Big{\rbrack}}
\newcommand{\rbc}{\Big{\rbrace}}
\newcommand{\lbc}{\Big{\lbrace}}

\title{Transverse Momentum Dependent Fragmenting Jet Functions with Applications to Quarkonium Production}

\author{Reggie Bain,}
\author{Yiannis Makris,}
\author{and Thomas Mehen}
\affiliation{Department of Physics, Duke University,\\
 Science Dr., Box 90305, Durham, NC 27708, USA}


\emailAdd{rab59@duke.edu}
\emailAdd{ym58@duke.edu}
\emailAdd{mehen@phy.duke.edu}

\abstract{We introduce the transverse momentum dependent fragmenting jet function (TMDFJF), which
appears in factorization theorems for cross sections for jets with  an identified hadron. These are functions of $z$, 
the hadron's longitudinal momentum fraction, and transverse momentum, $\veb{p}$, relative to the jet axis. In the 
framework of Soft-Collinear Effective Theory (SCET) we derive the
TMDFJF from both a factorized SCET cross section and the TMD fragmentation function defined in the literature. The
TMDFJFs are factorized into distinct collinear and soft-collinear modes by matching onto
SCET$_+$. As TMD calculations contain rapidity divergences, both the renormalization group (RG) and
rapidity renormalization group (RRG) must be used to provide resummed calculations with next-to-leading-logarithm prime
(NLL') accuracy. We apply our formalism to the production of $J/\psi$ within jets initiated by gluons. In this case
the TMDFJF can be calculated in terms of NRQCD (Non-relativistic  quantum chromodynamics) fragmentation functions. We 
find that when the $J/\psi$ carries a significant fraction of the jet energy, the $p_T$ and $z$ distributions differ 
for different NRQCD production mechanisms. Another observable with discriminating power is the average angle that the $J/\psi$ makes with the jet axis.
}

\begin{document}
\maketitle
\noindent


\section{Introduction}
In recent years, jet physics has played a prominent role at high energy colliders, particularly the Large Hadron Collider (LHC). 
Jets provide an opportunity to test our understanding of Quantum Chromodynamics (QCD) and appear in 
both Standard Model and beyond the Standard Model cross sections, making them important for searches of new physics as well.
Due to the enormous energies available at the LHC, top quarks, $W^\pm$, $Z^0$, and Higgs bosons are frequently 
produced with transverse momenta much greater than their mass, and studies of jet substructure have proved essential 
in identifying these highly boosted particles when they decay hadronically \cite{Altheimer:2013yza,Sapeta:2015gee}. 
For all these reasons, precision jet calculations have become increasingly important in particle physics. At the heart of
analytic calculations of jets are factorization theorems which separate jet cross sections into perturbative and
non-perturbative pieces. Non-perturbative functions such as parton distribution functions (PDFs), fragmentation
functions (FFs), and fragmenting jet functions (FJFs) offer ways to analytically probe the structure of the proton as well as the nature of hadronization.

FJFs were first introduced in Ref.~\cite{Procura:2009vm} within the framework of Soft-Collinear Effective Theory (SCET)~\cite{Bauer:2001yt,Bauer:2000ew,Bauer:2000yr,Bauer:2001ct}.
They appear in factorization theorems for cross sections for jets containing an identified hadron $h$ carrying a
fraction $z$ of the jet energy
($z=2E_h/\omega$, where $E_h$ is the hadron energy and $\omega=\sum_{i \in jet} p^-_{i}$). Ref.~\cite{Procura:2009vm}
also showed that one can construct a factorization theorem in
SCET for the differential cross-section $d \sigma^h / dz$ from the inclusive jet cross section by applying
a simple replacement rule,
\begin{equation}
  \label{eq:rule}
J_i(s,\mu)\rightarrow \frac{1}{2(2\pi)^3}{\cal{G}}_{i/h}(s,z,\mu) dz,
\end{equation}
where $J_i$ is the standard jet function and ${\cal{G}}_{i/h}$ is a FJF. In Eq.(\ref{eq:rule}), $s$ is the jet
invariant mass and $\mu$ is the renormalization scale.
Additionally in the limit $\Lambda_{\text{QCD}} \ll \sqrt{s}$ we can evaluate the FJF as a convolution of perturbative
short distance coefficients and the more commonly studied
fragmentation functions (FFs).
\begin{equation}
{\cal{G}}_{i/h}(s,z,\mu)= \sum_{j} \int_{z}^{1} \frac{dx}{x} \; {\cal{J}}_{i/j}(s,x,\mu) D_{j/h}\left(\frac{z}{x},\mu \right) + {\cal{O}}(\Lambda_{\text{QCD}}^2/s).
\end{equation}
Past studies of FJFs involve the jet invariant mass~\cite{Jain:2011xz,Procura:2009vm,Liu:2010ng} and were generalized to
angularities~\cite{Berger:2003iw} in Ref.~\cite{Bain:2016clc}. In Ref.~\cite{Procura:2011aq,Baumgart:2014upa,Chien:2015ctp}
FJFs independent of jet substructure observables were used to study the production of light quarks, heavy mesons, and quarkonia.
The properties of FJFs have also been studied
in Refs.~\cite{Jain:2011iu, Jain:2012uq, Bauer:2013bza, Dai:2016hzf, Kang:2016ehg, Ritzmann:2014mka, Kang:2016ioz, Metz:2016swz}.
In this work we extend FJFs to transverse momentum dependent distributions (TMDs). Recently, TMDs have been studied extensively
within and outside the framework of SCET~\cite{GarciaEchevarria:2011rb,Echevarria:2012pw,Echevarria:2014rua,Echevarria:2015usa,Echevarria:2016scs,Anselmino:2013lza,Procura:2014cba,Musch:2010ka,Chiu:2012ir}.
TMDs offer a new technology for the study of hadron substructure in hadron colliders (TMD parton distribution functions
(TMDPDFs)) and hadron production (TMD fragmentation functions (TMDFFs)). TMDPDFs have been used in SCET for studies of
Higgs production in the small transverse momentum limit at the
LHC~\cite{Neill:2015roa, Chiu:2012ir, Luebbert:2016itl,Becher:2012yn, Becher:2012uya,Ahrens:2008nc}. TMD fragmenting
jet functions (TMDFJF) depend on three kinematic variables: the jet energy, $\omega/2$, the fraction of this energy
carried by the identified hadron, $z$, and the hadron transverse momentum with respect to the axis of direction
of the original parton, $\veb{p}^h$.
The modes that give important contributions to the transverse momentum are
\begin{align}
\text{collinear-soft:} &\;\;\; p_{cs}^{\mu}\sim \omega(\lambda r,\lambda/r, \lambda),\;\;\;\; \lambda=p_{\perp}/\omega\nn \\
\text{collinear:}&\;\;\;p_{n}^{\mu} \sim \omega (\lambda^2,1,\lambda),
\end{align}
where collinear-soft modes are soft modes collinear to the direction of the jet axis first introduced in Ref.~\cite{Bauer:2011uc} and  $r \equiv \tan{\left(R/2\right)}$ for jet
cone size $R$. Similar modes are also studied in~\cite{Chien:2015cka}. To incorporate contributions from soft-collinear modes, we make use of the $\text{SCET}_+$ formalism. $\text{SCET}_+$ and other similar extensions of SCET
have been used to study processes with multiple well-separated scales and distinct phase space regions (e.g.~\cite{Bauer:2011uc,Pietrulewicz:2016nwo,Procura:2014cba,Chien:2015cka}).

Recent work~\cite{Baumgart:2014upa,Bain:2016clc} shows that jet substructure observables may be able to shed light on
outstanding puzzles in the production of quarkonia such as $J/\psi$ and $\Upsilon$. Our modern understanding of quarkonium
production comes from non-relativistic QCD (NRQCD)~\cite{Bodwin:1994jh}, an effective field theory that writes cross sections
and decay rates for bound states of heavy quarks as expansions in the strong coupling $\alpha_s(2m_c)$ and $v$, the relative
velocity of the quark-antiquark pair. NRQCD provides factorization theorems~\cite{Braaten:1993mp, Braaten:1993rw, Braaten:1994vv, Braaten:1996jt}
for cross sections in terms of perturbatively calculable short-distance
pieces multiplied by  non-perturbative long distance matrix elements (LDMEs). The perturbative piece describes the creation of
a heavy quark-antiquark pair in a given color and angular momentum state while the non-perturbative LDMEs
describe the hadronization of the heavy quark-antiquark pair into the physical quarkonium state. The different intermediate
color and angular momentum states of the pair define different NRQCD production mechanisms for quarkonia.

Ref.~\cite{Baumgart:2014upa} studied the dependence of the cross section for the production of $J/\psi$ within jets initiated by gluons
on $z$, the fraction of the jet's energy, $E_J$, carried by the $J/\psi$. The authors showed that the $z$ dependence is sensitive
to the underlying quarkonium production mechanism. Thus, simultaneously measuring the $z$ and $E_J$ dependence of the cross section for $J/\psi$ production within jets
provides a new and independent way of extracting the values of the LDMEs. Ref.~\cite{Bain:2016clc} extended these results
to $J/\psi$ production in $e^+e^-$ collisions where the angularity of the jet was probed. Ref.~\cite{Bain:2016clc} also found
that NLL' resummed analytic calculations of the $z$ distributions were quite different from those predicted by PYTHIA simulations.
The authors attributed this large discrepancy to an unrealistic modeling of the shower radiation from color-octet quarkonium
production mechanisms.

Intuitively, one might expect color-octet quark-antiquark intermediate states to radiate more gluons relative to color-singlet pairs.
This would result in $J/\psi$ produced with higher $p_{\perp}$ relative to the jet axis. Also, since different
color-octet production mechanisms have different FFs in NRQCD, FJFs should be able to distinguish
between the different color-octet production mechanisms. In addition to generalizing FJFs to TMD distributions, this paper
also shows that these TMDFJFs do in fact provide discriminating power between the different mechanisms.

In Section~\ref{sec:TMDFJF}, we give a definition of the TMDFJF and show how it emerges from definitions of TMDFFs in the literature. We then perform a matching calculation at next-to-leading order (NLO) onto SCET$_+$ and derive a result that is completely
factorized into hard, collinear, collinear-soft, and ultra-soft modes. We present a calculation of the matching
coefficients $\cJ_{ij}$  between the TMDFJF and the more traditionally studied FFs.
Additionally, we present a perturbative calculation of the corresponding collinear-soft function at NLO. In Section~\ref{sec:numerical},
we use renormalization group (RG) and rapidity renormalization group (RRG) techniques to resum logarithms to next-to-leading-log-prime (NLL')
accuracy. The TMDFJF formalism is applied to the production of $J/\psi$ in gluon jets where the FFs are calculated to LO in NRQCD.
We find that distributions in $p_{\perp}$  and $z$ as well as the average angle of
$J/\psi$ relative to the axis of the jet  can discriminate between the various NRQCD production mechanisms. Conclusions are given in section 4.
Appendix~\ref{sec:AppA} gives calculational details of the matching of the TMDFJF onto the FF, Appendix~\ref{sec:AppB} has an
alternative derivation of the TMDFJF from an SCET factorization theorem for a jet cross section, and Appendix~\ref{sec:AppC}
has details about the RG and RRG evolution.


\section{Transverse momentum dependent fragmenting jet function}
\label{sec:TMDFJF}
In this section we will present the definition of the TMDFJF, connecting it with definitions of TMDFFs from the literature. 
We first show the matching calculation of the TMDFJF onto SCET$_+$ and its factorization into pure
collinear, soft-collinear, and hard pieces. We then present perturbative calculations of the matching coefficients, ${\cal J}_{i/j}$, 
from matching the pure collinear function onto the FF as well as the one-loop expression for the soft-collinear function.
\subsection{Definition and factorization}
The operator definition of the quark  FF is given by~\cite{Collins:1981uw}:
\begin{equation}
  \label{eq:OPFF}
D_{q/h}(z,\mu)=\frac{1}{z} \sum_{X}\frac{1}{2 N_c} \delta(\omega-p^{-}_{X}-p^{-}_h)\; \Tr
\Big{\lbrack}\frac{\slashed{\bar{n}}}{2} \langle 0 \vert \psi(0)\vert X h \rangle \langle X h \vert
\bar{\psi}(0)\vert 0  \rangle  \Big{\rbrack} \Big{\vert}_{\veb{p}^{X}=-\veb{p}^h},
\end{equation}
where $\psi(x)$ is the quark field in QCD. The TMDFF is given by a similar expression but is unintegrated in the transverse components of the hadron momentum.
It is defined by~\cite{Aybat:2011zv}
\begin{equation}
   \label{eq:OPTMDFF}
D_{q/h}(\veb{p}^{h},z,\mu)=\frac{1}{z} \int \frac{d^{2} x_{\perp}}{(2 \pi)^2} \sum_{X}\frac{1}{2 N_c}
\delta(\omega-p^{-}_{X}-p^{-}_h)\; \Tr \Big{\lbrack}\frac{\slashed{\bar{n}}}{2} \langle 0 \vert
\psi(0,0,x_{\perp})\vert X h \rangle \langle X h \vert \bar{\psi}(0)\vert 0  \rangle  \Big{\rbrack},
\end{equation}
such that,
\begin{equation}
\int d^{2}\veb{p}^h \; D_{q/h}(\veb{p}^{h},z,\mu) = D_{q/h}(z,\mu).
\end{equation}
Here, $\veb{p}^{h}$ is the transverse momentum of the hadron $h$ with respect to the direction of the
original fragmenting quark. In order to identify the experimentally measured jet axis with the direction of the parton
initiating the jet, there needs to be a constraint that only ultrasoft
radiation is outside the jet. Alternative definitions of the TMDFF often involve the transverse momentum
measured with respect to different axes (e.g., the beam axis). In order to extend this concept to identified hadrons within jets we consider the collinear limit of Eq.(\ref{eq:OPTMDFF}) by matching onto SCET where now $z \equiv E_h/E_J$. This yields the operator definition of the TMDFJF
\begin{multline}
  \label{eq:OPFJF}
\hspace{-0.3cm}{\cal{G}}_{q/h}(\veb{p},z,\mu)=  \frac{1}{z} \sum_X \frac{1}{2 N_c} \delta(p^{-}_{Xh;r})\delta^{(2)}
(\veb{p}+\veb{p}^{X})
\Tr \Big{\lbrack} \frac{\slashed{\bar{n}}}{2} \langle 0 \vert \delta_{\omega,\overline{\cal{P}}}  \chi^{(0)}_{n} (0)\vert X h
\rangle \langle X h \vert \bar{\chi}^{(0)}_{n}(0)\vert 0  \rangle   \Big{\rbrack} ,
\end{multline}
where in the equation above the states $\vert X h \rangle$ corresponds to the a final state of collinear particles within a jet, in contrast with the the state $\vert X h \rangle$ in Eqs. (\ref{eq:OPFF}) and (\ref{eq:OPTMDFF}) which correspond to the inclusive case. The index $(0)$ indicates that the field has been decoupled from the ultra-soft gluons via BPS field redefinitions
\begin{equation}
  \chi_{n,\omega}^{(0)}(x)= Y_n^{\dag}(x)\chi_{n}(x) \quad \text{and} \quad A_n^{(0)}(x) = Y_n^{\dag}(x)A_n(x)Y_n(x),
\end{equation}
and $\chi_{n}\equiv  W_n^{\dag}\xi_{n}$
is defined in terms of the collinear quark fields of SCET and the ultrasoft and collinear Wilson lines are
\begin{equation}
Y_n^{\dag}(x) = \mathbb{P}\exp{\left( ig\int_0^{\infty}{ ds\; n \cdot A_{us}(x+sn) } \right)}
  \quad \text{and} \quad W_n(x) = \sum_{\text{perms}}{ \exp{\left( \frac{-g}{\bar{n}\cdot \cP} \bar{n}\cdot A_n{x}\right)} }.
\end{equation}
As we show in Appendix~\ref{sec:AppB},
the expression for the TMDFJF given in Eq.(\ref{eq:OPFJF}) is closely related to the FJF introduced in Ref.~\cite{Procura:2009vm}.

As discussed in the introduction, the TMDFJF receives contributions from two different modes, collinear and colllinear-soft or csoft .
In order to make the contribution of the csoft modes explicit, we now match our expression onto $\text{SCET}_{+}$,
\begin{multline}
{\cal{G}}_{q/h}(\veb{p},z,\mu)= C_+^{\dag}(\mu) C_{+}(\mu)  \frac{1}{z} \sum_X \frac{1}{2 N_c} \delta(p^{-}_{Xh;r})\delta^{(2)}  (\veb{p}+\veb{p}^{X})  \\
\times \Tr \Big{\lbrack} \frac{\slashed{\bar{n}}}{2} \langle 0 \vert \delta_{\omega,\overline{\cal{P}}} V_{n}^{\dag\;(0)}(0) \chi^{(0)}_{n} (0)\vert X h \rangle
\langle X h \vert \bar{\chi}^{(0)}_{n}(0) V_{n}^{(0)}(0)\vert 0  \rangle   \Big{\rbrack} ,
\end{multline}
where
\begin{equation}
  V_{n}^{(0)}(x)=\sum_{\text{perm}} \exp \left( \frac{-g}{\bar{n}\cdot {\cal{P}}} \bar{n}\cdot A_{n,\,cs}^{(0)}(x) \right),
\end{equation}
are Wilson lines of csoft fields (the csoft analogue of $W_n$) and $C_{+}(\mu)$ are SCET$_+$ matching coefficients.
In order to decouple the collinear fields $A_n^{(0)}$ and $\chi^{(0)}_n$ from the csoft gluons, we  now perform
field redefintions similar to those of the BPS procdure~\cite{Bauer:2011uc}
\begin{multline}
{\cal{G}}_{q/h}(\veb{p},z,\mu)= C_+^{\dag}(\mu) C_{+}(\mu)  \frac{1}{z} \sum_X \frac{1}{2 N_c} \delta(p^{-}_{Xh;r}) \delta^{(2)}  (\veb{p}+\veb{p}^{X})  \\ \times
 \Tr \Big{\lbrack} \frac{\slashed{\bar{n}}}{2} \langle 0 \vert  \delta_{\omega,\overline{\cal{P}}} V_{n}^{\dag\;(0)}(0) U_n^{(0)}(0)  \chi^{(0,0)}_{n} (0)\vert X h \rangle \\ \times
\langle X h \vert \bar{\chi}^{(0,0)}_{n}(0) U_{n}^{\dag \;(0)}(0) V_{n}^{(0)}(0) \vert 0  \rangle   \Big{\rbrack},
\end{multline}
where
\begin{equation}
  U_{n}^{\dag\,(0)}(x)=\mathbb{P}  \exp \left( i g \int_{0}^{\infty} ds \; n \cdot A^{(0)}_{n,\,cs}(ns+x)  \right),
  \end{equation}
and the superscript $(0,0)$ denotes that the corresponding fields are decoupled from both ultra-soft and collinear-soft modes.
Having factorized our operators, we now factorize the phase-space into collinear and collinear-soft Hilbert states.
\begin{equation}
\vert X h\rangle \; \rightarrow\; \vert X_n h\rangle \vert X_{cs}\rangle,
\end{equation}
\begin{equation}
\sum_{X} \; \rightarrow\; \sum_{X_n}\sum_{X_{cs}},
\end{equation}
\begin{equation}
\delta^{(2)}  (\veb{p}+\veb{p}^{X}) \;\rightarrow \; \delta^{(2)}  (\veb{p}+\veb{p}^{X_n}+\veb{p}^{X_{cs}}).
\end{equation}
This allows us to factorize the TMDFJF into three pieces 
\begin{equation}
{\cal{G}}_{q/h}(\veb{p},z,\mu)= H_+(\mu)\times \lb {\cal{D}}_{q/h}\otimes_{\perp}S_{C} \rb (\veb{p},z,\mu)\,,
\end{equation}
where $H_+$ is proportional to the square of the matching coefficient from ${\cal{G}}_{q/h}$ in SCET$_I$ to SCET$_+$, and ${\cal{D}}_{q/h}$  and $S_{C}$ are the  contributions collinear and the collinear-soft modes of SCET$_+$ to the TMDFJF, respectively. These are defined by
\begin{equation}
H_+ (\mu) = (2 \pi)^2 N_c \;C_+^{\dag}(\mu) C_+(\mu) \, ,
\end{equation}
\begin{multline}
{\cal{D}}_{q/h}(\veb{p}^{\cal{D}},z)\equiv\frac{1}{z} \sum_{X_{n}} \frac{1}{2 N_c} \delta(p^{-}_{Xh;r})\delta^{(2)}(p^{\perp}_{Xh;r})
\Tr \Big{\lbrack} \frac{\slashed{\bar{n}}}{2} \langle 0 \vert \delta_{\omega,\overline{\cal{P}}} \chi_{n} (0)\delta^{(2)}  ({\cal{P}}_{\perp}^{X_n}+\veb{p}^{\cal{D}})\vert X_n h \rangle \\ \times
\langle X_n h \vert \bar{\chi}_{n}(0) \vert 0  \rangle \Big{\rbrack},
\end{multline}
\begin{equation}
  \label{eq:soft}
 S_C(\veb{p}^S) \equiv \frac{1}{N_c} \sum_{X_{cs}}  \Tr \Big{\lbrack} \langle 0 \vert  V_{n}^{\dag}(0) U_n(0) \delta^{(2)} ({\cal{P}}_{\perp}+\veb{p}^{S})
 \vert X_{cs} \rangle \langle X_{cs}
   \vert U_{n}^{\dag}(0) V_{n}(0) \vert 0  \rangle \Big{\rbrack},
\end{equation}
where the $\Tr$ is over Dirac and color indices in $\cD_{q/h}$ and color indices in $S_C$. From now on, we drop the $(0)$ and $(0,0)$ superscripts since the different
collinear, soft-collinear, and ultra-soft modes are now factorized. We also employ the following shorthand for the convolution in the $\perp$ components
\begin{equation}
{\cal{D}}_{q/h} \otimes_{\perp} S_{C} (\veb{p}) = \int \frac{d^{2}\veb{p}'}{(2 \pi)^2}  \; \;{\cal{D}}_{q/h}(\veb{p}-\veb{p}') S_{C}(\veb{p}').
\end{equation}
Analogously for gluon fragmentation we have
\begin{align}
 {\cal{D}}_{g/h}(\veb{p},z,\mu)=& -g_{\mu\nu} \frac{1}{z} \sum_X \frac{\omega}{(d-2)(N_c^2-1)} \delta(p^{-}_{Xh;r}) \delta^{(2)}  (\veb{p}+\veb{p}^{X})\nn \\
&\times \langle 0 \vert  \delta_{\omega,\overline{\cal{P}}} {\cal{B}}^{\nu,a}_{n,\perp} (0)\delta^{(2)}  ({\cal{P}}_{\perp}^{X_n}+\veb{p}^{\cal{D}})\vert X h
\rangle \langle X h \vert {\cal{B}}^{\mu,a}_{n,\perp}(0)\vert 0  \rangle,
\end{align}
where the collinear gluon field is
\begin{equation}
  B_{n,\perp}^{\mu}(y) = \frac{1}{g}\left[ W_n^{\dagger}(y)iD_{n\perp}W_n(y)\right] \, ,
\end{equation}
and $iD_{n\perp}= \cP_{n\perp}^{\mu}+gA_{n\perp}^{\mu}$ is the standard $\perp$-collinear covariant derivative in SCET.

At this point, only the purely collinear term ${\cal{D}}_{i/h}$ contains information about the hadron $h$. The collinear-soft function ($S_C$) and the
hard function ($H_{+}$) are universal functions dependent on the fragmenting parton $i$ but not on the hadron $h$. Additionally, in the limit that
$\vert \veb{p} \vert \gg \Lambda_{\text{QCD}}$, we may use the operator product expansion to factorize ${\cal{D}}_{i/h}$ into short distance
coefficients and the more commonly studied FFs, $D_{j/h}$, via,
\begin{equation}
{\cal{D}}_{i/h}(\veb{p},z,\mu,\nu)= \int_{z}^{1} \frac{dx}{x}\;{\cal{J}}_{i/j}(\veb{p},x,\mu,\nu) D_{j/h}\left(\frac{z}{x},\mu\right)\;\;+\;\;{\cal{O}}
\left(\frac{\Lambda_{QCD}^2}{\vert\veb{p}\vert^2} \right),
\end{equation}
where ${\cal{J}}_{i/j}$ are the short distance coefficients that do not depend on the final hadron and can be calculated order by order in perturbation theory.

\begin{center}
\begin{figure}[h]
\centerline{\includegraphics[scale=0.57]{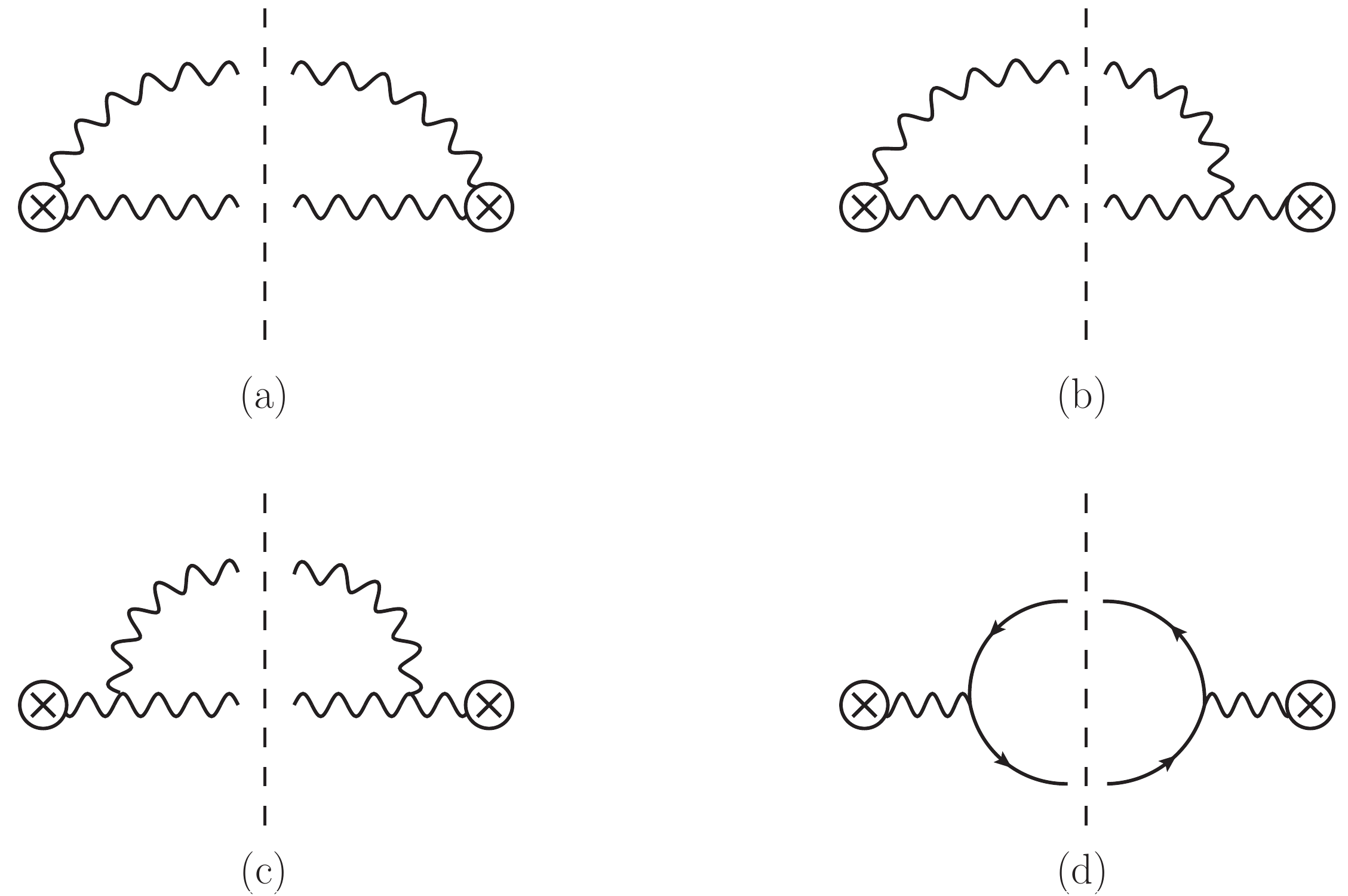}}
\caption{Feynman diagrams that give non-scaleless contributions to the gluon TMDFJF at NLO in $\alpha_s$. Diagram (b) also has a mirror image that is
	      not explicitly shown.}
\label{fig:gg}
\end{figure}
\end{center}
\begin{center}
\begin{figure}[h]
\centerline{\includegraphics[scale=0.135]{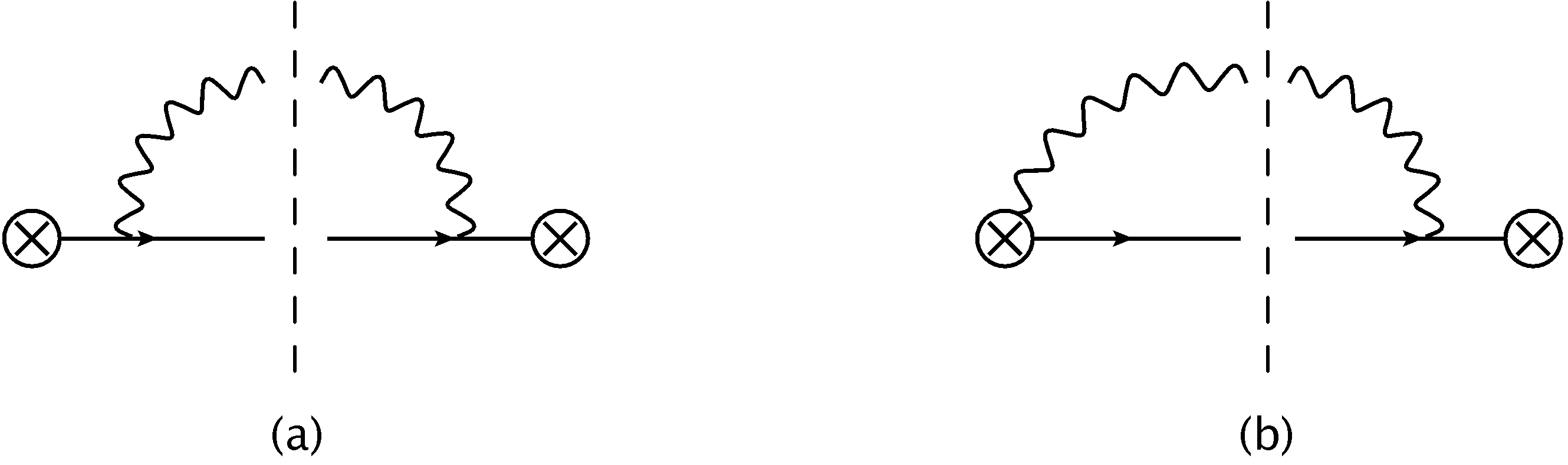}}
\caption{Associated non-scaleless diagrams that contribute to the quark TMDFJF at NLO. Again, Diagram (b) has a mirror image that is not explicitely drawn above.}
\label{fig:qq}
\end{figure}
\end{center}

\subsection{Perturbative results}
The $O(\alpha_s)$ diagrams contributing to the gluon and quark TMDFJFs
are shown in Figs.~\ref{fig:gg} and \ref{fig:qq}, respectively.
At NLO, the matching coefficients ${\cal{J}}_{i/j}$ are directly related to the matching coefficients ${\cal{I}}_{i/j}$ between TMDPDFs and
the more commonly studied PDFs calculated in Refs.~\cite{Procura:2014cba,Chiu:2012ir} by the substitution ${\cal I}_{i/j} \to {\cal J}_{j/i}$. See Appendix~\ref{sec:AppA} for additional
details of the matching calculation. Following Ref.~\cite{Chiu:2012ir}, a rapidity regulator is used to regulate rapidity divergences in the perturbative calculation.
This is implemented by first modifying the form of the collinear and collinear-soft Wilson lines
\begin{align}
\begin{split}
&W_n = \sum_{\text{perms}}{ \exp{\left( -\frac{g \;w^2}{\bar{n} \cdot \cP} \frac{\abs{\bar{n}\cdot \cP_g}^{-\eta}}{\nu^{-\eta}}
      \bar{n}\cdot A_n\right)} }\\
&V_n = \sum_{\text{perms}}{ \exp{\left( -\frac{g\; w}{\bar{n} \cdot \cP} \frac{\abs{\bar{n}\cdot \cP_g}^{-\eta/2}}{\nu^{-\eta/2}}
 \bar{n}\cdot A_{n,cs}\right)} },\\
\end{split}
\end{align}
with similar modifications to $U_n$. This introduces a regulator $\eta$, a bookkeeping parameter $w$, and a new dimensionful parameter
$\nu$. The dependence of our results on $\nu$ should of course cancel amongst the terms in our factorization theorem.
The renormalized results for the ${\cal{J}}_{i/j}$ in the $\overline{\text{MS}}$ scheme can be written,
\begin{multline}
  {\cal{J}}_{i/j}(\veb{p},z,\mu,\nu)= \delta_{ij}\delta(1-z)\delta^{(2)}(\veb{p})\\
  +\frac{\alpha_s T_{ij}}{\pi} \Big{\lbrace} \left( \delta_{ij}\delta(1-z)\ln\left(\frac{\omega^2}{\nu^2}\right)+\bar{P}_{ji}(z)\right) {\cal{L}}_0(\veb{p}^2,\mu^2)
  +c_{ij}(z)\delta^{(2)}(\veb{p}),
\Big{\rbrace},
\end{multline}
with
\begin{align}
&\bar{P}_{qq}(z)=P_{qq}(z)-\bar{\gamma}_q\delta(1-z)=(1+z^2){\cal{L}}_0(1-z), \nn \\
  &\bar{P}_{gq}(z)=P_{gq}(z)=\frac{1+(1-z)^2}{z}, \nn \\
  &\bar{P}_{qg}(z)=P_{qg}(z)=z^2+(1-z)^2, \nn \\
&\bar{P}_{gg}(z)=P_{gg}(z)-\bar{\gamma}_g\delta(1-z)=2\frac{(1-z+z^2)^2}{z}{\cal{L}}_0(1-z),
\end{align}
and
\begin{equation}
{c}_{qq}(z)=\frac{1-z}{2}, \;\;
{c}_{qg}(z)=\frac{z}{2}, \;\;
{c}_{gg}(z)=0, \;\;
{c}_{gq}(z)=z(1-z),
\end{equation}
where $T_{qq}=T_{qg}=C_F,\;T_{gg}=C_A,\;T_{gq}=T_F,\;\bar{\gamma}_q=3/2$ and  $\bar{\gamma}_g= \beta_0/(2C_A)$. For convenience we use the following
shorthand notation for the vector plus-distributions,
\begin{equation}
{\cal{L}}_n (\veb{p}^2,\mu^2)=\frac{1}{2 \pi \mu^2}{\cal{L}}_n \left(\frac{\veb{p}^2}{\mu^2}\right)=\frac{1}{2 \pi \mu^2}\left(\frac{\mu^2 }
{\veb{p}^2}\ln^{n}(\mu^2/\veb{p}^2) \right)_{+}.
\end{equation}
Performing the convolutions in the energy ratio parameter $z$ we get,
\begin{multline}
{\cal{D}}_{i/h}(\veb{p}^2,z,\mu,\nu)= D_{i/h}(z,\mu)\delta^{(2)}(\veb{p})+ \frac{\alpha_s}{\pi} \lbc \lb T_{ii} D_{i/h}(z,\mu)
\ln \left( \frac{\omega^2 (1-z)^2}{\nu^2} \right)\\
 +f^{i/h}_{P \otimes D} (z,\mu) \rb {\cal{L}}_0(\veb{p}^2,\mu^2) + f^{i/h}_{c \otimes D} (z,\mu) \delta^{(2)} (\veb{p})
\rbc,
\end{multline}
where
\begin{multline}
  f^{i/h}_{P \otimes D} (z,\mu)= \sum_{j} \lbc \delta_{ij} T_{ii} \int_{z}^{1} \frac{dx}{1-x} \; \lb p_{i}(x)
  D_{i/h}\left(\frac{z}{x},\mu\right)-2 D_{i/h}\left(z,\mu \right) \rb \\
  + (1-\delta_{ij})T_{ij} \int_{z}^{1} \frac{dx}{x} \; P_{ji}(x) D_{j/h}\left(\frac{z}{x},\mu\right) \rbc,
\end{multline}
with $p_q(x)=(1+x^2)/x$,  $p_g(x)=2(1-x+x^2)^2/x^2$ and
\begin{equation}
f^{i/h}_{c \otimes D} (z,\mu)= \sum_{j} T_{ij} \int_{z}^{1} \frac{dx}{x} \; c_{ij}(x) D_{j/h}\left(\frac{z}{x},\mu\right),
\end{equation}

\begin{center}
\begin{figure}[h]
\centerline{\includegraphics[scale=0.57]{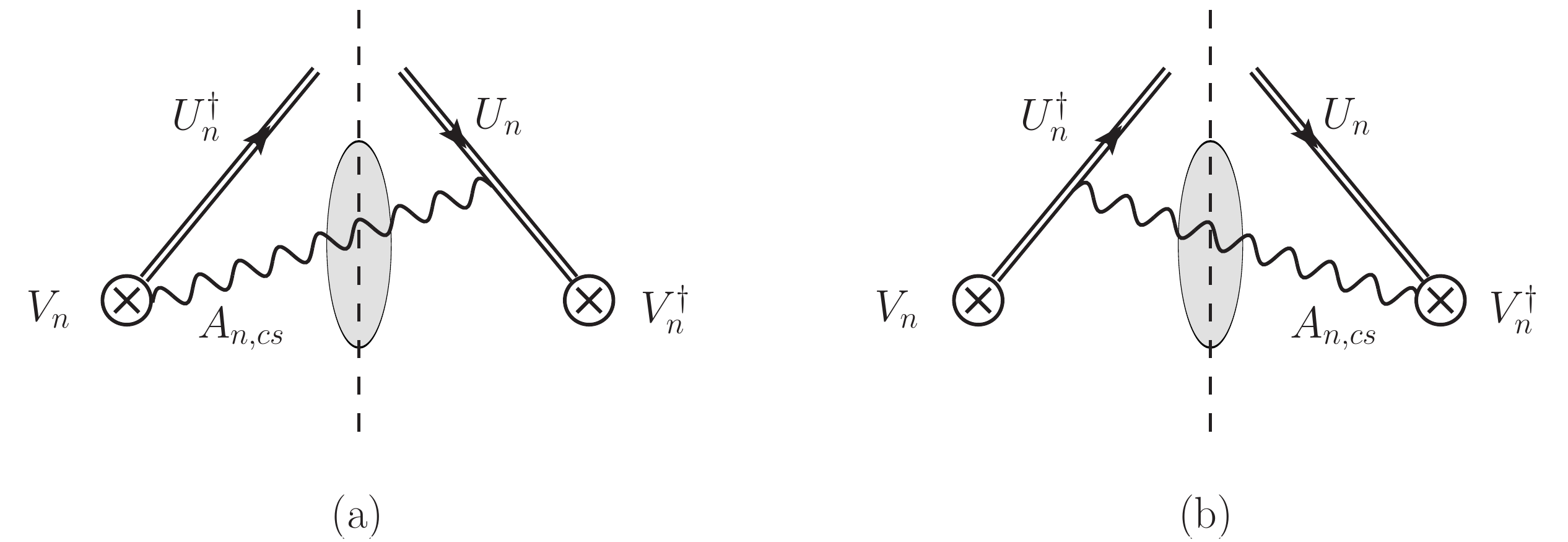}}
\caption{Real gloun emission diagrams that contribute to the collinear-soft function $S_C^{i}(\veb{p},z,\mu,\nu)$ at ${\cal{O}}(\alpha_s)$.
The gluons passing through the shaded oval indicate they are contained within the phase-space of the jet.}
\label{fig:soft}
\end{figure}
\end{center}
At NLO, the collinear-soft function, defined by Eq.~(\ref{eq:soft}), receives contributions from the two diagrams shown in Fig.~\ref{fig:soft}.
The real gluon is contained within a jet defined by a cone or $k_T$-type jet algorithm with cone size parameter $R$. A global soft funciton of similar form
has been calculated at NLO in Ref.~\cite{Chiu:2012ir} and at NNLO in Ref.~\cite{Luebbert:2016itl} in studies of Higgs $p_{T}$ spectrum. The two diagrams in Fig.~\ref{fig:soft}
yield identical contributions and thier sum is given by,
\begin{align}
S_{C}^{i, B(1)}(\veb{p})&=+g^2 w^2 \left(\frac{e^{\gamma_E}\mu^2}{4 \pi}\right)^{\epsilon} \nu^{\eta}\;C_{i} \int\frac{dk^+ dk^- d^{d-2}k_{\perp}}{2(2\pi)^{d-1}}\;
\frac{2}{k^+ (k^-)^{1+\eta}} \delta(k^2)\delta^{(2)}(\veb{k}+\veb{p})\;\Theta_{\text{alg}} \nn \\
& = +\frac{\alpha_s w^{2} C_{i} }{\pi} \frac{e^{\gamma_E\epsilon}}{\Gamma(1-\epsilon)} \left( \frac{\nu r}{\mu}\right)^{ \eta} \frac{1}{\eta} \frac{1}{2 \pi \mu^2}
\left(\frac{\mu^2}{\veb{p}^2} \right)^{1+\epsilon+\eta/2},
\end{align}
where $\Theta_{\text{alg}}$ defines the jet algorithm, $r \equiv \tan(\mathrm{R}/2)$, and $C_q=C_F,\;C_g=C_A$. After an expansion in $\eta$ followed by an
expansion in $\epsilon$ and summing both diagrams we get,
\begin{align}
S_{C}^{i,B} (\veb{p},\mu,\nu)= & \delta^{(2)}(\veb{p})+ \frac{\alpha_s w^2 C_{i}}{\pi}
\Big{\lbrace} \frac{2}{\eta}\left(-\frac{1}{2\epsilon}\delta^{(2)}(\veb{p})+{\cal{L}}_0(\veb{p}^2,\mu^2) \right) \nn \\
&+ \delta^{(2)}(\veb{p})\left(\frac{1}{2\epsilon^2}+\frac{1}{2\epsilon}\ln \left(\frac{\mu^2}{r^2\nu^2}\right) \right)-{\cal{L}}_0(\veb{p}^2,\mu^2)
\ln \left(\frac{\mu^2}{r^2\nu^2}\right)+{\cal{L}}_1(\veb{p}^2,\mu^2)\nn \nn \\
&-\frac{\pi^2}{24}\delta^{(2)} (\veb{p})
\Big{\rbrace},
\end{align}
The renormalized result (where we have now set $w\to 1$)in the $\overline{\text{MS}}$ scheme is thus
\begin{equation}
S_{C}^{i,R} (\veb{p},\mu,\nu)=  \delta^{(2)}(\veb{p})- \frac{\alpha_s C_{i}}{\pi}
\Big{\lbrace} {\cal{L}}_0(\veb{p}^2,\mu^2)\ln \left(\frac{\mu^2}{r^2\nu^2}\right)-{\cal{L}}_1(\veb{p}^2,\mu^2)+\frac{\pi^2}{24}\delta^{(2)} (\veb{p})
\Big{\rbrace}.
\end{equation}
While in general this expression receives contributions from virtual gluon emission diagrams at NLO, these diagrams yield scaleless integrals when
using this particular set of regulators. Thus virtual diagrams are neglected and all singularities from these real emission diagrams are interpreted as UV divergences.
We also verified, using a set of regulators where such virtual gluons give non-zero contributions, that the result is identical.\footnote{In order to verify that
all IR divergences do indeed cancel, we used a gluon mass, rapidity regulator, and dimensional regulator where diagrams with virtual
gluons give non-scaleless contributions.} Note if  pure dimensional  regularization is used for ultraviolet and infrared divergences then $H_+=(2\pi)^2 N_c$ as discussed in~Ref.~\cite{Procura:2014cba}.

\section{Numerical Results}
\label{sec:numerical}
\subsection{Renormalization Group (RG) and Rapidity Renormalization Group (RRG)}
Individual diagrams for the collinear-soft function $S_C$ and the matching coefficients $\cJ_{i/j}$ suffer from infra-red (IR), ultra-violet (UV) and rapidity divergences (RD).
We use dimensional regularization and a rapidity regulator (as introduced and developed in Ref.~\cite{Chiu:2011qc,Chiu:2012ir}) to regulate these divergences.
IR divergences in the collinear-soft function cancel when summing over all diagrams. In the matching coefficients ${\cal{J}}_{i/j}$,
IR divergences cancel in the matching  of the collinear functions $\cD_{i/h}$ onto traditional FFs, $D_{j/h}$.
The remaining poles (UV and rapidity), are removed by renormalization. In addition to the scale $\mu$ introduced by dimensional regularization
our use of a rapidity regulator requires the introduction of an additional scale, $\nu$. With this scale are associated rapidity renormalization group (RRG) equations
which can be used to resum  large logarithms by evolving each function from its canonical scale to a common scale.
Bare and renormalized quantities are related through the following convolution with the renormalization factor $Z$,
\begin{equation}
F^{B}(\veb{p})=Z_{F}(\veb{p},\mu,\nu) \otimes_{\perp} F^{R}(\veb{p},\mu,\nu),
\end{equation}
where $F$ can be either ${\cal{D}}_{i/h}$ or $S_C^i$ and satisfies the following RG and RRG equations,
\begin{align}
& \frac{d}{d \ln \mu}F^{R}(\veb{p},\mu,\nu) = \gamma^{F}_{\mu} (\mu,\nu) \times F^{R} (\veb{p},\mu,\nu) \nn \\
& \frac{d}{d \ln \nu}F^{R}(\veb{p},\mu,\nu) = \gamma^{F}_{\nu} (\veb{p},\mu,\nu) \otimes_{\perp} F^{R} (\veb{p},\mu,\nu).
\end{align}
Here $\gamma_{\mu}^F$ and $\gamma_{\nu}^{F}$ are the anomalous dimensions associated to RG and RRG respectively and are defined by,
\begin{align}
  \label{eq:AD}
   \lb (2 \pi)^2 \delta^{(2)}(\veb{p}) \rb  \times \gamma_{\mu}^{F} (\mu,\nu) &=- Z_{F}^{-1}(\veb{p},\mu,\nu) \otimes_{\perp} \frac{d}{d \ln \mu}Z_F(\veb{p},\mu,\nu) \nn \\
   \gamma_{\nu}^{F}(\veb{p},\mu,\nu) & =- Z_{F}^{-1}(\veb{p},\mu,\nu) \otimes_{\perp} \frac{d}{d \ln \nu}Z_F(\veb{p},\mu,\nu).
  \end{align}
For the renormalization factors we find,
\begin{align}
  Z^{\cal{D}} (\veb{p},\mu,\nu)=&(2\pi)^2 \delta^{(2)}(\veb{p})+(4 \pi)\alpha_s w^2 C_F
\Big{\lbrace}
-\frac{2}{\eta}\left(-\frac{1}{2 \epsilon}\delta^{(2)}(\veb{p})+{\cal{L}}_0(\veb{p}^2,\mu^2) \right)\nn\\
  &+\frac{1}{2\epsilon}\left(\ln\left(\frac{\nu^2}{\omega^2}\right)+\bar{\gamma}_i \right)\delta^{(2)}(\veb{p})
\Big{\rbrace}
\end{align}
\begin{align}
  Z^{S_C} (\veb{p},\mu,\nu)=&(2\pi)^2 \delta^{(2)}(\veb{p})+(4 \pi)\alpha_s w^2 C_F
\Big{\lbrace}
+\frac{2}{\eta}\left(-\frac{1}{2\epsilon}\delta^{(2)}(\veb{p})+{\cal{L}}_0(\veb{p}^2,\mu^2) \right)\nn\\
& +\frac{1}{2\epsilon}\left(\ln \left(\frac{\mu^2}{r^2\nu^2}\right)+\frac{1}{\epsilon} \right)\delta^{(2)}(\veb{p}),
\Big{\rbrace}
\end{align}
The $\mu$ anomalous dimensions are found using Eq. (\ref{eq:AD}),
\begin{equation}
\gamma_{\mu}^{\cal{D}}(\nu)=\frac{\alpha_s C_i}{\pi} \left(\ln\left(\frac{\nu^2}{\omega^2}\right)+\bar{\gamma}_i \right)
\end{equation}
\begin{equation}
\gamma_{\mu}^{S_C}(\nu)=\frac{\alpha_s C_i}{\pi} \ln\left(\frac{\mu^2}{r^2\nu^2}\right),
\end{equation}
For the $\nu$ anomalous dimensions, our bookkeeping parameter $w$ plays an analogous role to the coupling $g$ for the case of the
$\mu$ anomalous dimension, although $w$ itself is not a coupling, such that,
\begin{equation}
  \nu \pf{}{\nu}w = -\frac{\eta}{2}w,
\end{equation}
thus yielding
\begin{equation}
\label{eq:nuAD1}
\gamma_{\nu}^{\cal{D}}(p_{\perp},\mu)=-(8 \pi) \alpha_s C_i\; {\cal{L}}_0(\veb{p},\mu^2)
\end{equation}
\begin{equation}
\label{eq:nuAD2}
\gamma_{\nu}^{S_C}(p_{\perp},\mu)=+(8 \pi) \alpha_s C_i\; {\cal{L}}_0(\veb{p},\mu^2).
\end{equation}
The anomalous dimensions satisfy
\begin{equation}
\gamma_{\mu}^{\cal{D}}(\nu)+\gamma_{\mu}^{S_C}(\nu)=\gamma_{\mu}^{J}=\frac{\alpha_s C_i}{\pi} \left(\ln\left(\frac{\mu^2}{r^2 \omega^2}\right)+\bar{\gamma}_i \right),
\end{equation}
where $\gamma_{J}$ is the anomalous dimension of the unmeasured quark jet function~\cite{Ellis:2010rwa} and
\begin{equation}
\gamma_{\nu}^{\cal{D}}(\veb{p},\mu)+\gamma_{\nu}^{S}(\veb{p},\mu)=0.
\end{equation}
In order to resum our results to NLL' accuracy we evolve the purely collinear function and the collinear-soft function from their characteristic scales where logarithms are
minimized to common scales in $\mu$ and $\nu$ using the RG and RRG respectively. To perform the evolution, we first solve the Fourier transforms of both the RRG and RG equations.
We then perform the evolution using the RG and RRG before finally performing the inverse Fourier transform.
The simplest resummation procedure is, in this case, to first evolve our collinear-soft function in RRG space and choose the common scale to be $\nu=\nu_{\cal{D}}$.
We then evolve both functions in RG space to the common scale $\mu=\omega r$. Notice that $S_C$ and $\cD$ have the same characteristic renormalization
scale $\mu_{S_C}=\mu_{\cal{D}}\equiv \mu_C$. The equivalence of the virtualities of the soft and collinear modes is a defining feature of $\mathrm{SCET_{II}}$.

To make the interpretation of our plots easier, we study the quantity ${\cal{G}}_{i/h}(p_{\perp},z,\mu)$ which is related to the TMDFJF by the change
of variables from vector transverse momenta ($\veb{p}$) to the amplitude ($p_{\perp}=\vert \veb{p}\vert$). Performing the evolutions described above we find,
\begin{multline}
  \label{eq:finalev}
  {\cal{G}}_{i/h}(p_{\perp},z,\mu)= (2 \pi)^2 \; p_{\perp} \int_{0}^{\infty} db \;b J_{0}(bp_{\perp}){\cal{U}}_{S_C}(\mu,\mu_{S_C},m_{S_C}){\cal{U}}_{\cal{D}}(\mu,\mu_{\cal{D}},1)
  \\ \times {\cal{V}}_{S_C}(b,\mu_{S_C},\nu_{\cal{D}},\nu_{S_C}) {\cal{FT}} \lb {\cal{D}}_{i/h}(\veb{p},z,\mu_{\cal{D}},\nu_{\cal{D}})
  \otimes_{\perp}S_C^{i}(\veb{p},\mu_{S_C},\nu_{S_C}) \rb,
\end{multline}
where $\boldsymbol{b}$ is the Fourier conjugate variable of $\veb{p}$, $J_0$ is a Bessel function of the first kind,
\begin{equation}
 {\cal{U}}_F (\mu,\mu_0,m_{F})=\exp\left(K_F(\mu,\mu_0)\right) \left( \frac{\mu_0}{m_F} \right)^{\omega_F(\mu,\mu_0)},
\end{equation}
\begin{equation}
  \text{and}\;\;{\cal{V}}_F(b,\mu,\nu,\nu_{0})= \left( \frac{\mu}{\mu_C(b)} \right)^{\eta_F(\mu,\nu,\nu_0)} \quad \text{where} \quad \mu_C(b)=2 \exp(-\gamma_E)/b,
\end{equation}
are the evolution kernels resulting from solving the RG and RRG equations respectively.
The pure collinear term ${\cal{D}}_{i/h}$ in Eq.(\ref{eq:finalev}) involves the convolution of
the perturbatively calculated short distance coefficients and the standard fragmentation
functions evolved from their canonical scale to the canonical scale of the collinear term in
momentum space, $\mu=p_{\perp}$. The form of the fragmentation functions is fixed during the
Fourier transforms in Eq.(\ref{eq:finalev}). The scales $\mu_F$, $\nu_F$ and $m_F$ for each of
the functions are given in Table~\ref{tb:scales} and more details of the RG and RRG evolution
are provided in Appendix~\ref{sec:AppC}.
\begin{table}[h]
\begin{center}
\begin{tabular}{|c|c|c|c|}
  \hline
Function $(F)$     & RG scale ($\mu_F$) &  RRG scale ($\nu_F$) & $m_F$             \\
\hline{}
 ${\cal{D}}_{i/h}$ & $\mu_C(b)$         & $\omega$              & n.a.              \\
\hline
$S^{i}_C$          & $\mu_C(b)$         & $\mu_C(b)/r$          & $ \nu r $            \\
\hline
\end{tabular}
\end{center}
\caption{Characteristic scales of the different functions in the factorization theorem.}
\label{tb:scales}
\end{table}
\subsection{Applications to quarkonium production}
In this section we apply our TMDFJF formalism to the production of quarkonium in jets. We will focus on $J/\psi$ production within 
jets initiated by gluons, though our results can be easily generalized to $\Upsilon$ or other quarkonia and jets initiated by quarks. 
For $J/\psi$ production the leading production mechanism in the NRQCD $v$ expansion is ${}^3S_1^{[1]}$, where $^{2S+1}L_{J}^{[1,8]}$ indicates
 the color and angular momentum quantum numbers of the $c\bar{c}$ produced in the short-distance process. This mechanism scales as $v^3$,
whereas the leading color-octet mechanisms, ${}^3S_1^{[8]}, {}^1S_0^{[8]},$ and ${}^3P_J^{[8]}$, scale as $v^7$. 
Table \ref{tb:LDME} shows this scaling along with numerical values of the corresponding LDME extracted from the fits  in Ref.~\cite{Butenschoen:2011yh, Butenschoen:2012qr} (which we use below).
The extracted LDME are consistent with the $v^4$ suppression expected from NRQCD. As was done for the FJF's
in Ref.~\cite{Bain:2016clc} we use the leading order NRQCD~\cite{Bodwin:1994jh} FFs for gluon fragmentation to $J/\psi$
for each of the four mechanisms. 
In the $\alpha_s$ expansion the leading order contribution to gluon fragmentation to $J/\psi$ via the ${}^3S_1^{[1]}$ mechanism scales as $\alpha_s(2m_c)^3$, while for ${}^1S_0^{[8]}$ and ${}^3P_J^{[8]}$ the leading contribution scales as $\alpha_s(2 m_c)^2$ and for the ${}^3S_1^{[8]}$ mechanism the fragmentation function scales as 
$\alpha_s(2m_c)$. Thus for gluon fragmentation the $v^4$ suppression of color-octet mechanisms is compensated for by fewer powers of $\alpha_s$ and all four contributions 
are roughly the same size. Our goal is to see if the $z$ and $p_\perp$ dependence of the TMDFJF can discriminate between these competing mechanisms.

\begin{center}
\begin{figure}[h]
\centerline{\includegraphics[scale=0.62]{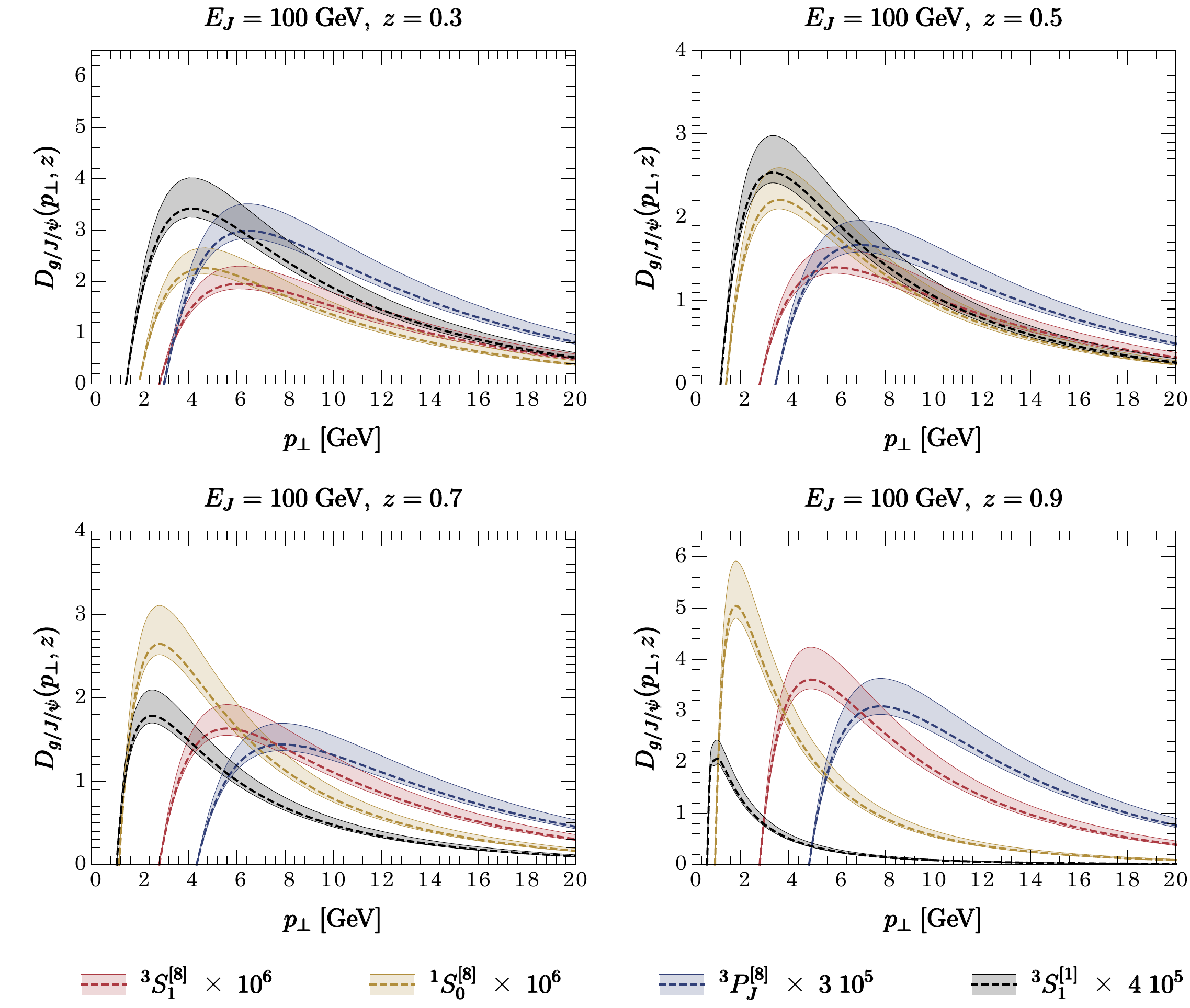}}
\caption{The TMDFJF as a function of the $p_{\perp}$ of the $J/\psi$ for the ${}^3S_1^{[1]}, {}^3S_1^{[8]}, {}^1S_0^{[8]}, {}^3P_J^{[8]}$ production mechanisms where the
 for jet energies $E_J=100\; \text{GeV}$. Theoretical uncertainties are calculated by varying the renormalization
 scales by factors of  $1/2$ and $2$.}
\label{fig:plotpT1}
\end{figure}
\end{center}
\begin{table}[h]
\begin{center}
\begin{tabular}{|c|c|c|c|c|}
\hline
$\langle\mathcal{O}^{J/\psi}(^3S_1^{[1]}) \rangle$  & $\langle \mathcal{O}^{J/\psi}(^3S_1^{[8]}) \rangle $ & $\langle \mathcal{O}^{J/\psi}(^1S_0^{[8]}) \rangle$ & $\langle \mathcal{O}^{J/\psi}(^3P_0^{[8]}) \rangle $ \\
\hline
$\sim v^3 $  & $\sim v^7 $ & $ \sim v^7  $& $\sim v^7  $\\
\hline{}
1.32 GeV$^3$& 2.24 $\times 10^{-3}$ GeV$^3$ & $4.97 \times 10^{-2}$ GeV$^3$ & $-1.61\times 10^{-2}$GeV$^5$\\
\hline
\end{tabular}
\end{center}
\caption{LDMEs for NRQCD production mechanisms. Here $v$ is the relative velocity of $c \bar{c}$ pair. For charmonium $v \sim 0.3$. For the numerical result we use central values taken from global fits in Refs.~\cite{Butenschoen:2011yh, Butenschoen:2012qr}. }
\label{tb:LDME}
\end{table}

\begin{center}
\begin{figure}[h]
\centerline{\includegraphics[scale=0.62]{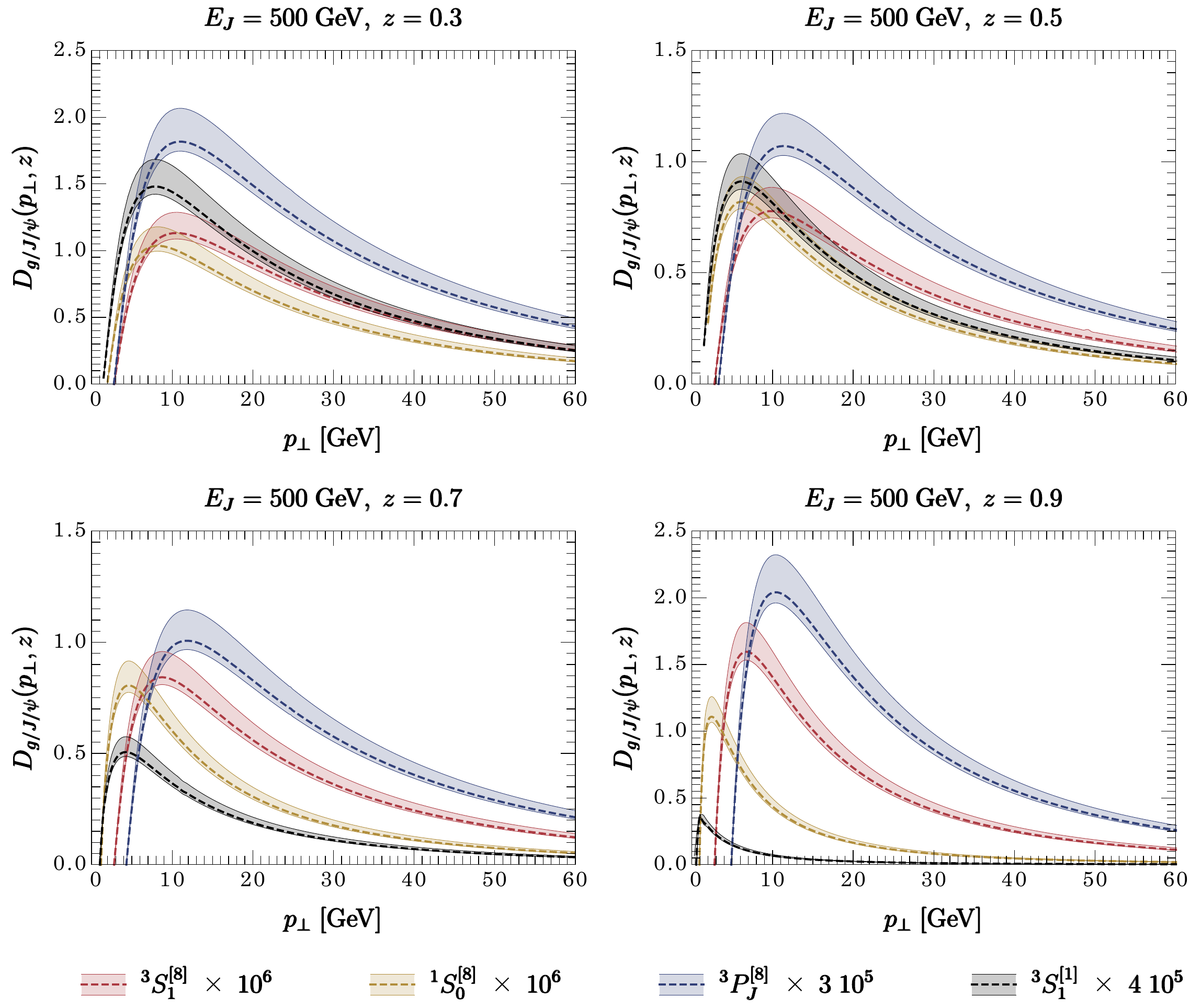}}
\caption{The TMDFJF as a function of the $p_{\perp}$ of the $J/\psi$ for the $^3S_1^{[1]}, {}^3S_1^{[8]}, {}^1S_0^{[8]}, ^3P_J^{[8]}$ production mechanisms where the
 for jet energies $E_J= 500 \,\text{GeV}$. Theoretical uncertainties are calculated by varying the renormalization
 scales by factors of $1/2$ and $2$.}
\label{fig:plotpT2}
\end{figure}
\end{center}

The TMDFJF as a function of $p_\perp$ for fixed $z$, for $z=0.3,0.5,0.7,$ and $0.9$, are shown Figs.~\ref{fig:plotpT1} and~\ref{fig:plotpT2}, for jet energies of 
100 ${\rm GeV}$ and 500 GeV, respectively. In order to make it easier to view all distributions simultaneously, we have rescaled the ${}^3S_1^{[8]}$, ${}^1S_0^{[8]}$, ${}^3P_J^{[8]}$,and ${}^3S_1^{[1]}$ distributions, by factors of $10^6$, $10^6$, $3.0 \, 10^5$ and $4.0 \, 10^5$, respectively. The same rescaling factor is used in all eight plots in Figs.~\ref{fig:plotpT1} and~\ref{fig:plotpT2}, and 
theoretical uncertainties are calculated by varying the RRG and RG scales $\nu_{S_C},\;\nu_{\cD},\;\text{and}\;\mu$ by a factor of $2$ and $1/2$.
The central dashed lines in the figures correspond to the scale choices $\nu=\nu_{\cD}=\omega$ and  $\mu=\omega r$. Though we plot our distributions in the 
range $0 < p_\perp < 20$ GeV, it is important that to keep in mind that our calculations are only reliable for $p_\perp \geq 2 m_c = 3$ GeV.

These plots show that the TMDFJF does in fact provide discriminating power amongst the four mechanisms. For $z = 0.3$, all four distributions look similar 
for both $E_J =100 $ GeV and 500 GeV. The distributions peak at roughly the same location and they have same slope for large $p_\perp$. For $z\geq 0.5$, the color-singlet ${}^3S_1^{[1]}$ mechanism and the color-octet ${}^1S_0^{[8]}$ mechanism peak at lower values of $p_\perp$ and  fall more steeply 
with $p_\perp$ than the ${}^3S_1^{[8]}$ and ${}^3P_J^{[8]}$ color-octet mechanisms. The ${}^3P_J^{[8]}$ mechanism has the peculiar feature that in order to obtain a positive FF we need to have a negative LDME, as is found in the fits of Refs.~\cite{Butenschoen:2011yh, Butenschoen:2012qr}.  The peaks in the $p_\perp$ distribution for the ${}^3S_1^{[1]}$ and ${}^1S_0^{[8]}$ mechanisms are at very low $p_\perp$ where perturbation theory is not reliable. On the other hand, the peaks of the 
${}^3S_1^{[8]}$ and ${}^3P_J^{[8]}$ distributions are at larger values of $p_\perp \sim 6-8$ GeV where perturbation theory can be trusted. The  ${}^3P_J^{[8]}$ gives a slightly harder $p_\perp$ distribution than  ${}^3S_1^{[8]}$ mechanism, and both are significantly harder than the other mechanisms.

\begin{figure}[t]
\centerline{\includegraphics[scale=0.62]{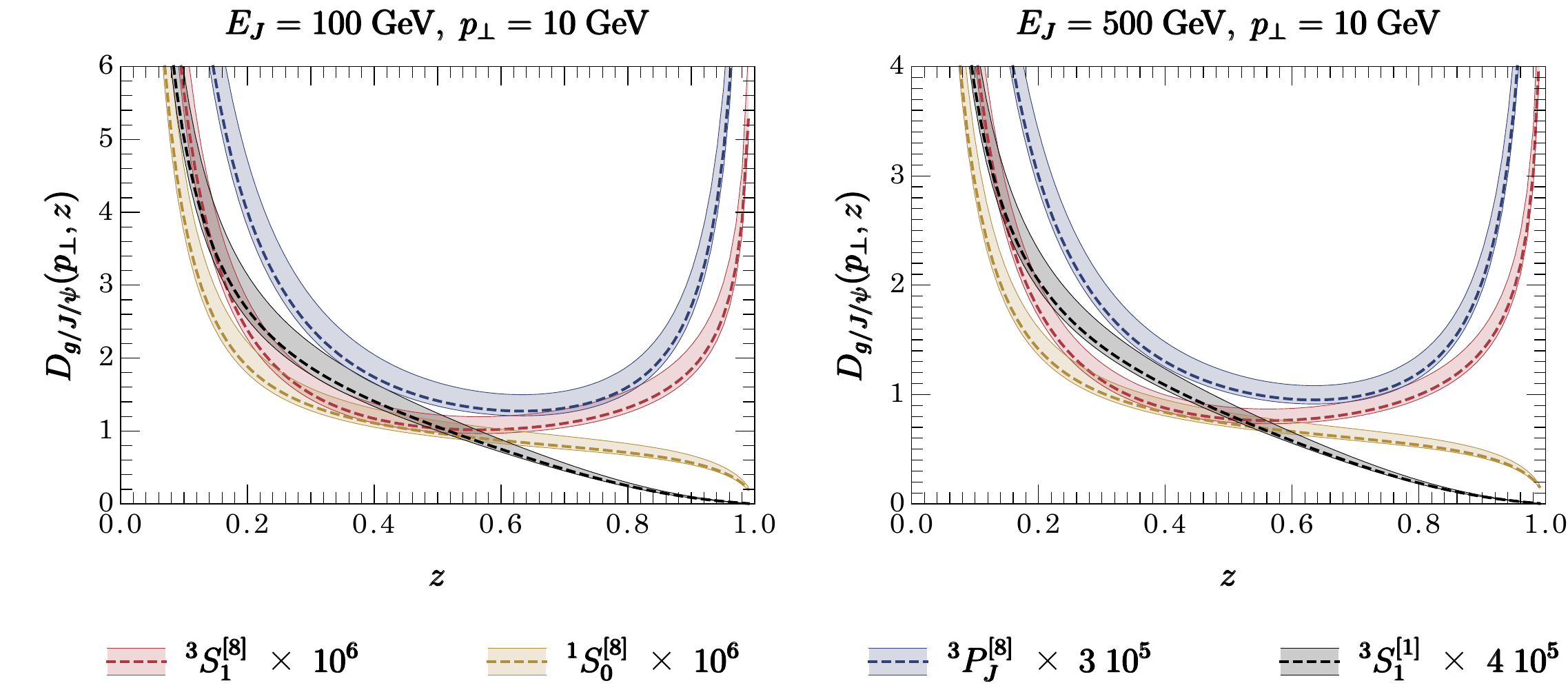}}
\caption{The TMDFJF as a function of the $z$ of the $J/\psi$ for the $^3S_1^{[1]}, {}^3S_1^{[8]}, {}^1S_0^{[8]}, ^3P_J^{[8]}$ production mechanisms,
with $p_\perp = 10$ GeV for $E_J = 100$ and $500$ GeV.  Theoretical uncertainties are calculated by varying the renormalization
 scales by factors of $1/2$ and $2$. }
\label{fig:plotz}
\end{figure}

It is interesting to study the dependence of the TMDFJF as a function of $z$ with $p_\perp$ fixed to be a perturbative scale. In Fig.~\ref{fig:plotz} we plot the TMDFJF as 
a function of $z$ for $p_\perp = 10$ GeV for jets with energy $E_J=$ 100 and 500 GeV. Large logarithms and shape function effects will affect these distributions 
in both the $z \to 0$ and $z \to 1$ limits, but our calculations should be reliable for intermediate values of $z$.  While for $z<0.5$ the distributions have similar shapes, 
in the range $0.5 < z < 0.9$, the shapes of all four mechanisms are different. The $z$ dependence of the TMDFJF for fixed $p_\perp$ can be used to differentiate between the NRQCD production mechanisms. 
 

\begin{figure}[h]
\centerline{\includegraphics[scale=0.62]{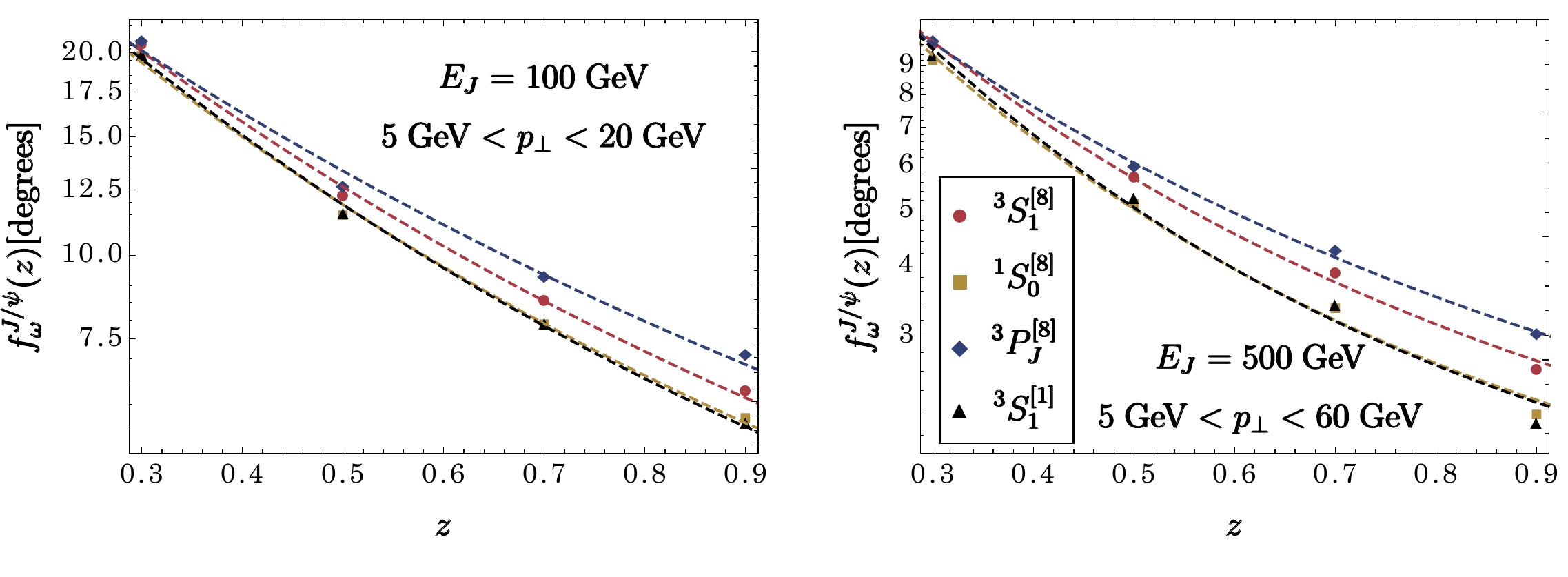}}
\caption{The  function $f^{J/\psi}_{\omega}(z)$ (as defined in the text) as a function of $z$ relative to the jet axis for each NRQCD production mechanism where the jet has $E_J=\omega/2=100$ GeV(left)  and $500\, \text{GeV}$ (right). The $J/\psi$  is restricted to have $p_{\perp}\in\left[5,\,20\right]\,\text{GeV}$ in the 100 GeV jet 
and $p_{\perp}\in\left[5,\,60\right]\,\text{GeV}$ in the 500 GeV jet.}
\label{fig:plotang}
\end{figure}

The TMDFJF formalism also allows us to calculate the angle at which $J/\psi$ are produced relative to the jet axis. The average production angle for the $J/\psi$ is given by 
\begin{equation}\label{eq:theta}
  \langle \theta \rangle (z) = \frac{\int \theta d\theta  (d\sigma/d\theta dz)}{  \int d\theta  (d\sigma/d\theta dz)  }.
\end{equation}
Using the small angle approximation the differential cross section can be written as
\begin{equation}
  \frac{d\sigma}{d\theta dz}=\int d p_{\perp}\; \delta \left( \theta - \frac{2 p_{\perp}}{z \omega} \right) \frac{d\sigma}{dp_{\perp} dz}.
\end{equation}
Substituting this into Eq.~\ref{eq:theta}  yields
\begin{equation}\label{eq:ratio}
  \langle \theta \rangle (z) =\frac{2\;\int dp_{\perp} p_{\perp} (d\sigma/dp_{\perp} dz)}{z \omega\;\int dp_{\perp} (d\sigma/dp_{\perp} dz) }.
\end{equation}
As discussed in Appendix~\ref{sec:AppB}, the cross section $d\sigma/d\theta dz $ can be factorized into hard, soft and collinear terms in SCET. 
In general the hard and soft contributions will not cancel because there is a sum over partonic channels in both the numerator and denominator of Eq.~\ref{eq:ratio}.
However, they will if gluon fragmentation dominates production, then the expression above  can be written as
\begin{equation}
  \langle \theta \rangle (z) \sim \frac{2 \int dp_{\perp} \; p_{\perp} {\cal{G}}_{g/h}(p_{\perp},z,\mu)}{z \omega \int dp_{\perp} \; {\cal{G}}_{g/h}(p_{\perp},z,\mu)} \equiv f^{h}_{\omega}(z),
  \end{equation}
where ${\cal{G}}_{g/h}(p_{\perp},z,\mu)$ is the gluon TMDFJF. Fig.~\ref{fig:plotang} the function $f_{\omega}^{J/\psi}(z)$ is plotted at points $z=0.3,0.5,0.7,$ and $0.9$ for $\omega=2E_J =200\,\text{GeV}$ and 
$\,1\,\text{TeV}$ for $J/\psi$ with $p_\perp \in [5,20]$ GeV and  $p_\perp \in [5,60]$ GeV, respectively.
As was done earlier we have fixed the scale $\mu = \omega r$. Note the typical angles are small enough that the small angle approximation is justified. The  dashed lines in figure show the results of a fit to the functional form, $C_0 \exp(-z \,C_1)$, the values of $C_0$ and  $C_1$ for each mechanism at each energy are shown in Table~\ref{tb:fit}.
Again we see that differences between the various NRQCD mechanisms become more pronounced as $z$ increases.
This shows that the average angle does in fact yield some discriminating power between the octet mechanisms. In particular the slope on the semilog plot, which is determined by the parameter $C_1$ in Table~\ref{tb:fit}, differs by as much as 20\% between the various NRQCD mechansims for $E_J = 100$ GeV and and as much as 40\% for $E_J =$ 500 GeV.
Note however that $^1S_0^{[8]}$ and  $^3S_1^{[1]}$ give very similar predictions for this observable.

\section{Conclusions}
In this paper we introduce the transverse momentum dependent  fragmenting jet function (TMDFJF) in the framework of SCET and  show how it is related to the
previously introduced TMDFFs and fragmenting jet functions (FJFs). TMDFJFs describe the transverse as well as longitudinal momentum  distribution of an identified hadron within
a jet. TMDFJFs evolve with the renormalization group (RG) scale $\mu$ and obey RG equations similar to jet functions.
Using $\text{SCET}_+$ we show that this new distribution can be further factorized into soft and purely collinear terms. The purely collinear factor can be written
as a convolution of perturbatively calculable short distance coefficients and the standard FFs, where the soft factor is given by a vacuum matrix
element of product of Wilson lines. This factorization introduces rapidity divergences that are regulated with the rapidity regulator. We check that at NLO the regulator dependance vanishes in the final product. Associated with rapidity divergences are rapidity renormalization group (RRG) equations. 
By evolving the  collinear and soft terms separately using  the RG and RRG  equations all orders resummation of large logarithms in the TMDFJF can be performed.

As an example we implement this formalism for the case of quarkonium production. In the case of quarkonia the TMDFJF can be calculated in terms of the 
NRQCD  FFs which are perturbatively calculable at the scale $2 m_Q$. For the gluon TMDFJF for $J/\psi$, we study the $p_\perp$ and $z$
dependence predicted by the four production mechanisms: $^3S_1^{[1]}, {}^3S_1^{[8]}, {}^1S_0^{[8]}$, and ${}^3P_J^{[8]}$. We use the leading order (in $\alpha_S$) NRQCD FF for each of these mechanisms, and the RG and RRG equations are used to calculate the TMDFJFs to next-to-leading-logarithmic-prime (NLL') accuracy. We find that the $z$ dependence 
(for fixed $p_\perp$) is different for all four mechanisms. We also find that the dependence on $p_\perp$ and the average angle of the $J/\psi$ relative to the jet axis can discriminate between the various NRQCD production mechanisms.

\begin{table}
\label{tb:fit}
\begin{minipage}[h]{0.6 \linewidth}
 \centering
 \begin{tabular}{| c | c | c |}
\hline
\multicolumn{3}{|c|}{$E_J= 100\,\text{GeV}$}\\
\hline{}
 $^{2S+1}L_{J}^{[1,8]}$  &  $C_0$  &  $C_1$   \\
\hline
 $^3S_1^{[1]}$  &  $3.92 $  &  $ 0.92 $   \\
 $^3S_1^{[8]}$  &  $3.86 $  &  $ 0.84 $   \\
 $^1S_0^{[8]}$  &  $3.88 $  &  $ 0.90 $   \\
 $^3P_J^{[8]}$  &  $3.75 $  &  $ 0.74 $   \\
 \hline
 \end{tabular}
\end{minipage}
\begin{minipage}[b]{0.31\linewidth}
  \centering
\begin{tabular}{| c | c | c |}
\hline
\multicolumn{3}{|c|}{$E_J = 500\,\text{GeV}$}\\
\hline{}
 $^{2S+1}L_{J}^{[1,8]}$  &  $C_0$  &  $C_1$ \\
\hline
 $^3S_1^{[1]}$  &  $3.75 $  &  $1.68 $  \\
 $^3S_1^{[8]}$  &  $3.48 $  &  $1.39 $  \\
 $^1S_0^{[8]}$  &  $3.66 $  &  $1.64 $  \\
 $^3P_J^{[8]}$  &  $3.28 $  &  $1.20 $  \\
 \hline
\end{tabular}
 \end{minipage}
\caption{Results of fits of $\log{\left(f_{\omega}(z)\right)}$ shown in Fig.~\ref{fig:plotang} to the function $C_0 \exp(-z\, C_1)$.}
\end{table}

\acknowledgments
The authors would like to thank Aneesh Manohar for his helpful suggestions on the manuscript. TM and YM are supported in part by the Director, Office of Science, Office of
Nuclear Physics, of the U.S. Department of Energy under grant numbers DE-FG02-05ER41368. RB is supported by
a National Science Foundation Graduate Research Fellowship under Grant No. 3380012.

\appendix

\section{Matching calculation}
\label{sec:AppA}
In this Appendix we provide details for the evaluation of the matching coefficients, ${\cal J}_{i/j}$. 
From the sum of diagrams in Figs.~\ref{fig:qq}a) and \ref{fig:qq}b) we get:
\begin{align}
  {\cal{D}}_{q/q}^{B(1)} (\veb{p},z,\mu,\nu)= &\frac{\alpha_s w^2 C_F}{\pi} \frac{e^{\gamma_E\epsilon}}{\Gamma(1-\epsilon)}
  \left(\frac{\nu}{\omega}\right)^{ \eta} \frac{1}{2 \pi \mu^2}\left(\frac{\mu^2}{\veb{p}^2}\right)^{1+\epsilon} \nn \\&
  \times \lbc 2z \left(\frac{1}{1-z}\right)^{1+\eta} + (1-\epsilon)(1-z)\rbc \nn\\
= &  \frac{\alpha_s w^2 C_F}{\pi}
\Big{\lbrace}
\Big{\lbrack}-\frac{2}{\eta}\left(-\frac{1}{2\epsilon}\delta^{(2)}(\veb{p})+{\cal{L}}_0(\veb{p}^2,\mu^2) \right) \nn \\& +\frac{1}{2\epsilon}
\left(\ln\left(\frac{\nu^2}{\omega^2}\right)+\frac{3}{2} \right)\delta^{(2)}(\veb{p})  \Big{\rbrack}\delta(1-z)
- \frac{1}{2\epsilon} P_{qq}(z)\delta^{(2)}(\veb{p})\nn \\& + \left(-\delta(1-z)\ln\left(\frac{\nu^2}{\omega^2}\right)+\bar{P}_{qq}(z)\right)
{\cal{L}}_0(\veb{p}^2,\mu^2)+c_{qq}(z)\delta^{(2)}(\veb{p})\Big{\rbrace} \nn \\
 &+{\cal{O}}(\eta,\epsilon),
\end{align}
where we define $c_{qq}(z)=(1-z)/2$. The superscripts $B$ and $R$ denote bare and renormalized quantities, respectively, and the superscript $(1)$ indicates that 
this is the $O(\alpha_S)$ contribution. The NLO matching coefficient is given by
\begin{equation}
{\cal{J}}_{q/q}^{R(1)}(\veb{p},z,\mu)={\cal D}_{q/q}^{R(1)}(\veb{p},z,\mu)-D_{q/q}^{R(1)}(z,\mu)\delta^{(2)}(\veb{p}),
\end{equation}
where
\begin{equation}
D_{q/q}^{R(1)}(z)=-\frac{\alpha_s C_F}{\pi}P_{qq}(z)\frac{1}{2\epsilon}.
\end{equation}
The $1/\epsilon$ pole appearing in the FF is interpreted as an infrared divergence. Although for extracting the renormalized matching coefficients ${\cal{J}}_{i/j}$ we can ignore scaleless integrals and interpret the finite terms as the renormalized result to that particular order,
 here we are interested in the origin of the  poles since this will allow us to extract the anomalous dimensions. Performing the matching we get:
\begin{align}
{\cal{J}}_{q/q}^{R}(\veb{p},z,\mu,\nu)=&\delta^{(2)}(\veb{p})\delta(1-z)+\frac{\alpha_s C_F}{\pi}
\Big{\lbrace} \left(\delta(1-z)\ln\left(\frac{\omega^2}{\nu^2}\right)+\bar{P}_{qq}(z)\right) {\cal{L}}_0(\veb{p}^2,\mu^2)\nn\\&+c_{qq}(z)\delta^{(2)}(\veb{p})
\Big{\rbrace}.
\end{align}
For the coefficient ${\cal{J}}_{q/g}$ we simply perform the replacement  $z \rightarrow (1-z)$ and drop $\delta(z)$ and plus-distributions since these
functions are always integrated for values of $z$ greater than zero. Thus
\begin{equation}
{\cal{J}}_{q/g}^{R}(\veb{p},z,\mu,\nu)=\frac{\alpha_s C_F}{\pi}\Big{\lbrace}\bar{P}_{gq}(z) {\cal{L}}_0(\veb{p}^2,\mu^2)+c_{qg}(z)
\delta^{(2)}(\veb{p}) \Big{\rbrace},
\end{equation}
where $c_{qg}(z)=c_{qq}(1-z)=z/2$. For the gluon splitting we get
\begin{align}
  {\cal{D}}_{g/g}^{B(1)}(\veb{p},z,\mu,\nu)= & \frac{\alpha_s C_A w^2}{\pi} \frac{\mathrm{e}^{\epsilon \gamma_E}}{\Gamma(1-\epsilon)}
  \left(\frac{\nu}{\omega}\right)^\eta \frac{1}{2 \pi \mu^2}\left(\frac{\mu^2}{\veb{p}^2}\right)^{1+\epsilon}\nn\\&
  \times 2\; \Big{\lbrack} \frac{z}{(1-z)^{1+\eta}}+\frac{(1-z)}{z}+z(1-z) \Big{\rbrack}.
\end{align}
Expanding in $\eta$ and $\epsilon$ we have
\begin{align}
{\cal{D}}_{g/g}^{B(1)}(\veb{p},z,\mu,\nu)=&\frac{\alpha_s C_A w^2}{\pi}   \Big{\lbrack} -
\frac{1}{2 \epsilon}\delta^{(2)}(\veb{p})+{\cal{L}}_0(\veb{p}^2,\mu^2)\Big{\rbrack} \nn \\
&\times  \Big{\lbrack}-\frac{2}{\eta}\delta(1-z)-\ln \left(\frac{\nu^2}{\omega^2}\right)\delta(1-z)+\bar{P}_{gg}(z)\Big{\rbrack} \nn \\
=& \frac{\alpha_s C_A w^2}{\pi} \Big{\lbrace}\Big{\lbrack}-\frac{2}{\eta}\left(-\frac{1}{2\epsilon}\delta^{(2)}(\veb{p})+{\cal{L}}_0(\veb{p}^2,\mu^2) \right) \nn \\&
+\frac{1}{2\epsilon}\left( \ln \left(\frac{\nu^2}{\omega^2}\right) +\frac{1}{2} \beta_0\right)\delta^{(2)}(\veb{p})  \Big{\rbrack}\delta(1-z)\nn \\
& - \frac{1}{2\epsilon} P_{gg}(z)\delta^{(2)}(\veb{p}) + \left(-\delta(1-z)\ln\left(\frac{\nu^2}{\omega^2}\right)+\bar{P}_{gg}(z)\right) {\cal{L}}_0(\veb{p}^2,\mu^2) \Big{\rbrace},
\end{align}
and since the corresponding FF is given by:
\begin{equation}
D_{g/g}^{R}(z)=\delta(1-z)-\frac{\alpha_s C_A}{\pi}P_{gg}(z)\frac{1}{2\epsilon}+{\cal{O}}(\alpha_s^2),
\end{equation}
where the $1/\epsilon$ pole is an infrared divergence, we have
\begin{equation}
{\cal{J}}_{g/g}^{R}(\veb{p},z,\mu,\nu)=\delta^{(2)}(\veb{p})\delta(1-z)+\frac{\alpha_s C_A}{\pi} \left(\delta(1-z)\ln\left(\frac{\omega^2}{\nu^2}\right)+\bar{P}_{gg}(z)\right) {\cal{L}}_0(\veb{p}^2,\mu^2).
\end{equation}
A similar calculation yields the kernel ${\cal{J}}_{g/q}$,
\begin{align}
{\cal D}_{g/q}^{B(1)}(\veb{p},z,\mu,\nu)&= \frac{\alpha_s T_F w^2}{\pi} \frac{\mathrm{e}^{\epsilon \gamma_E}}{\Gamma(2-\epsilon)}  \frac{1}{2 \pi \mu^2}\left(\frac{\mu^2}{\veb{p}^2}\right)^{1+\epsilon}\times \left(\bar{P}_{qg}(z)-\epsilon\right) \nn \\
&= \frac{\alpha_s T_F w^2}{\pi}\Big{\lbrace}
-\frac{1}{2 \epsilon} \bar{P}_{qg}(z) \delta^{(2)}(\veb{p})+ {\cal{L}}_0(\veb{p}^2,\mu^2)\bar{P}_{qg}(z)+c_{gq}(z)\delta^{(2)}(\veb{p})
\Big{\rbrace},
\end{align}
where $c_{gq}(z)=z(1-z)$. Performing the matching and since the corresponding FF is
\begin{equation}
D_{g/q}^{R}(z)= -\frac{\alpha_s T_F}{\pi}P_{qg}(z)\frac{1}{2\epsilon}+{\cal{O}}(\alpha_s^2),
\end{equation}
where again the $1/\epsilon$ pole is an infrared divergence, we get
\begin{equation}
{\cal{J}}_{g/q}^{R}(\veb{p},z,\mu,\nu)=\delta^{(2)}(\veb{p})\delta(1-z)+\frac{\alpha_s T_F}{\pi} \Big{\lbrace}
 {\cal{L}}_0(\veb{p}^2,\mu^2)\bar{P}_{qg}(z)+c_{gq}(z)\delta^{(2)}(\veb{p})
\Big{\rbrace}.
\end{equation}

\section{Factorization Theorems in SCET}
\label{sec:AppB}
Much like the standard FJFs, TMDFJFs appear in factorization theorems for cross-sections that are differential in $z$,
the fraction of a jet initiating parton's energy carried by an identified hadron, and $\veb{p}$, the transverse
momenta of the hadron measured from the parton's momentum. It is shown in Ref.~\cite{Ellis:2010rwa} that the cross-section for the production
of two jets in electron-positron annihilation can be written as,
\begin{equation}
  \label{eq:crosssec}
d\sigma = d\sigma^{(0)} H_2(\mu) \times S_{\Lambda}(\mu) \times  J^{q}_{n}(\omega,\mu) \times  J^{\bar{q}}_{\bar{n}}(\omega,\mu) \, ,
\end{equation}
where $d\sigma^{(0)}$ is the Born cross section, $H_2(\mu)$ is the hard function resulting from matching a 2-jet operator in full QCD onto the corresponding SCET operators, 
$S_{\Lambda}(\mu)$ is a soft function that describes soft scale cross-talk between the jets and the soft out-of-jet radiation
is constrained via $E_{\text{out}} < \Lambda$, and  $J_{n}(\omega,\mu)$ is a jet function that describes collinear radiation within a
jet in the $\hat{n}$ direction that has energy $E_{J}=\omega/2$ (here $\omega=E_{\text{cm}}$). The jet function can be defined
in SCET as
\begin{equation}
J^{q}_n(\omega,\mu)=\int \frac{dk^{+}}{2 \pi} \int d^{4} x \; \exp(i k^+ x^-/2) \frac{1}{N_{C}} \text{Tr}
 \left[ \frac{\slashed{\bar{n}}}{2} \langle 0 \vert \delta_{\omega,\overline{\cal{P}}}\; \delta_{0,{\cal{P}}_{\perp}} \chi_{n}(x) \bar{\chi}_{n}(0) \vert 0 \rangle  \right].
\end{equation}
To study jets with identified hadrons, we insert the following expression for the identity
\begin{equation}
\boldsymbol{1}=\sum_{X} \vert X \rangle \langle X \vert = \sum_{X} \sum_{h \in {\cal{H}}_i} \int \frac{dz d^2\veb{p}^h}{2(2\pi)^3} \vert X h(z,\veb{p}^h) \rangle \langle X h(z,\veb{p}^h) \vert
\end{equation}
\begin{multline}
  J^{q}_n(\omega,\mu)= \sum_{h\in {\cal{H}}_i} \int \frac{dz d^2 \veb{p}}{2 (2 \pi)^3}\int \frac{dk^{+}}{2 \pi} \int d^{4} x \; \exp(i k^+ x^-/2) \frac{1}{N_{C}} \\
  \times \sum_{X} \text{Tr}
\lb \frac{\slashed{\bar{n}}}{2} \langle 0 \vert \delta_{\omega,\overline{\cal{P}}} \; \delta_{0,{\cal{P}}_{\perp}}\chi_{n}(x) \vert X h(z,\veb{p}) \rangle
  \langle X h(z,\veb{p}) \vert \bar{\chi}_{n}(0) \vert 0 \rangle  \rb.
\end{multline}
where $h$ is an identified hadron within the jet. Performing the integration over $x$, which is the Fourier conjugate
of the residual momenta, and the residual $k^+$ yields
\begin{equation}
J^{q}_n(\omega,\mu)= \sum_{h \in {\cal{H}}_i} \int z dz d^2\veb{p} \; {\cal{G}}_{q/h}(\veb{p},z,\mu).
\end{equation}
Insetrting this back to Eq.(\ref{eq:crosssec}) we have
\begin{equation}
  d\sigma =\sum_{h \in {\cal{H}}_i} \int z dz d^2\veb{p} \;  d\sigma^{(0)} H_2(\mu) \times S_{\Lambda}(\mu) \times  {\cal{G}}_{q/h}(\veb{p},z,\mu) \times J^{\bar{q}}_{\bar{n}}(\omega,\mu) .\,
\end{equation}
which directly implies 
\begin{equation}
  \frac{d \sigma^{i/h}}{dz d^2\veb{p}} = d\sigma^{(0)} H_2(\mu) \times  S_{\Lambda}(\mu) \times   {\cal{G}}_{q/h}(\veb{p},z,\mu) \times J^{\bar{q}}_{\bar{n}}(\omega,\mu)  + {\cal{O}}\left( \frac{
  \Lambda}{E_{J}}, \frac{\Lambda_{\text{QCD}}^2}{p_{\perp}^2} \right).
\end{equation}
This suggests a rather powerful rule (already known to be true for the standard FJFs) for constructing the factorization theorem in SCET with identified 
hadron with measured transverse momenta :
\begin{equation}
   \frac{d \sigma^{i/h}}{dz d^2\veb{p}} =d\sigma \lb J^{i}(\omega,\mu) \rightarrow  {\cal{G}}_{i/h}(\veb{p},z,\mu)\rb 
  \end{equation}
\section{Solving the RG and RRG Equations}
\label{sec:AppC}
\subsection{RRG Evolution}
The RRG equation in momentum space for a renormalized function $F^R$ is given by
\begin{equation}
  \label{eq:RRGConv}
  \nu \frac{d}{d\nu} F^R(\veb{p},\mu,\mu/\nu) = \gamma_{\nu}^F(\veb{p},\mu,\nu) \otimes_{\perp}F^R(\veb{p},\mu,\mu/\nu),
\end{equation}
where the anomalous dimension can be written in the following generic form,
\begin{equation}
  \gamma_{\nu}^{F}(\veb{p},\mu,\nu)=\Gamma_{\nu}^{F}[\alpha_s] {\cal{L}}_0(\veb{p}^2,\mu^2)+\gamma_{\nu}^{F}[\alpha_s] \delta^{(2)}(\veb{p}),
\end{equation}
where
\begin{equation}
  \delta^{(2)}(\veb{p}) = \frac{1}{\pi}\delta(\veb{p}^2).
\end{equation}
The cusp and non-cusp parts of the anomalous dimension are listed in Table~\ref{tb:AD}.
Taking the Fourier transform of Eq. (\ref{eq:RRGConv}) yields,
\begin{equation}
  \label{eq:fsRRG}
  \frac{d}{d \ln \nu} \tilde{F}(b,\mu,\nu) =\tilde{\gamma}_{\nu}^{F}(b,\mu,\nu)  \tilde{F}(b,\mu,\nu),
\end{equation}
where the Fourier conjugate of $\veb{p}$ is $\boldsymbol{b}$ where $\abs{\boldsymbol{b}}=b$ and using the form of the anomalous dimensions
in Eq. (\ref{eq:nuAD1},\ref{eq:nuAD2}) gives that,
\begin{equation}
  \tilde{\gamma}_{\nu}^{F}(b,\mu,\nu) =-\frac{\Gamma_{\nu}^{F}[\alpha_s]}{(2 \pi)^2} \ln \left(\frac{\mu}{\mu_C (b)}  \right)+\frac{\gamma_{\nu}^{F}[\alpha_s]}{(2 \pi)^2},
\end{equation}
where $\mu_C(b)=2e^{-\gamma_E}/b$. Integrating  Eq. (\ref{eq:fsRRG}) yields
\begin{equation}
 \tilde{F}(b,\mu,\nu)= \tilde{F}(b,\mu,\nu_0) {\cal{V}}_F (b,\mu,\nu,\nu_0),
\end{equation}
where
\begin{equation}
  {\cal{V}}_F (b,\mu,\nu,\nu_0)= \exp \lb G_F(\mu,\nu,\nu_0) \rb\left( \frac{\mu}{\mu_C} \right)^{\eta_F(\mu,\nu,\nu_0)},
\end{equation}
with
\begin{equation}
  G_F(\mu,\nu,\nu_0)=\frac{\gamma_{\nu}^{F} [\alpha_s]}{(2 \pi)^2} \ln \left( \frac{\nu}{\nu_0} \right)\;\;\;\;\; \text{and}\;\;\;\;\; \eta_F(\mu,\nu,\nu_0)=-\frac{\Gamma_{\nu}^{F} [\alpha_s]}{(2 \pi)^2} \ln \left( \frac{\nu}{\nu_0} \right).
\end{equation}

\begin{table}[h]
\begin{center}
\begin{tabular}{|c|c|c|c|c|}
  \hline
Function $(F)$     & $ \Gamma_{\nu}^F$             & $\gamma_{\nu}^F  $    &$ \Gamma_{F}^0$  & $\gamma_{F}^0  $        \\
\hline
 ${\cal{D}}_{i/h}$ & $-(8 \pi)\alpha_s C_i+{\cal{O}}(\alpha_s^2)$     & ${\cal{O}}(\alpha_s^2)$    &  $ 0$  & $ 4C_i (\ln(\nu^2/\omega^2)+\bar{\gamma}_i)$  \\
\hline
$S^{i}_C$          & $ (8 \pi)\alpha_s C_i+{\cal{O}}(\alpha_s^2)$     & ${\cal{O}}(\alpha_s^2)$    &  $ 4C_i$   & $ 0$    \\
\hline
\end{tabular}
\end{center}
\caption{Values of the cusp and non-cup parts of the anomalous dimensions for the collinear and collinear-soft functions.}
\label{tb:AD}
\end{table}

\subsection{RG Evolution}
Evolution in $\mu$ begins with the following RG equation
\begin{equation}
\frac{d}{d \ln \mu}F^{R}(\veb{p},\mu,\nu) = \gamma^{F}_{\mu} (\mu,\nu) \times F^{R} (\veb{p},\mu,\nu),
\end{equation}
where the anomalous dimension can be written in the generic form
\begin{equation}
  \gamma_{\mu}^F(\mu)=\Gamma_{\mu}^F [\alpha] \ln \left( \frac{\mu^2}{m_F^2} \right)+\gamma_{\mu}^{F}[\alpha].
\end{equation}
The coefficient $\Gamma_{\mu}^F [\alpha_s] $ is proportional to the cusp
anomalous dimension, $\Gamma_{\mathrm{cusp}}[\alpha_s]$, which can be expanded in $\alpha_s$
\begin{equation}
\Gamma_{\mathrm{cusp}}(\alpha_s)=\sum_{n=0}^{\infty} \left( \frac{\alpha_s}{4 \pi}\right)^{1+n} \Gamma_c^n,
\end{equation}
and $\Gamma_{\mu}^F =(\Gamma_F^0/\Gamma_c^0)\Gamma_{\mathrm{cusp}}$. The non-cusp part, $\gamma_{\mu}^{F}[\alpha_s]$, has a similar expansion
\begin{equation}
\gamma_{\mu}^{F}[\alpha_s]=\sum_{i=0}^{\infty} \left( \frac{\alpha_s}{4 \pi}\right)^{1+i} \gamma_{F}^i.
\end{equation}
The solultion to the RGE is thus given by
\begin{equation}
F^R(\veb{p},\mu,\nu)= F^R(\veb{p},\mu_0,\nu) {\cal{U}}_F(\mu,\mu_0,m_F)\, ,
\end{equation}
where again
\begin{equation}
  {\cal{U}}_F(\mu,\mu_0,m_F)= \exp \left( K_F (\mu, \mu_0) \right) \left( \frac{\mu_0}{m_F} \right) ^{\omega_F(\mu, \mu_0)}
  \end{equation}
and the exponents $K_F$ and $\omega_F$ are given in terms of the anomalous dimension,
\begin{align}
K_F(\mu, \mu_0) &= 2 \int_{\alpha (\mu)}^{\alpha(\mu_0)} \frac{d \alpha'}{\beta(\alpha')} \Gamma_F (\alpha') \int_{\alpha(\mu_0)}^{\alpha'}
\frac{d \alpha''}{\beta(\alpha'')} +\int_{\alpha (\mu)}^{\alpha(\mu_0)} \frac{d \alpha'}{\beta(\alpha')} \gamma_F (\alpha') ,\\
\omega_F(\mu, \mu_0) &= 2 \int_{\alpha (\mu)}^{\alpha(\mu_0)} \frac{d \alpha'}{\beta(\alpha')} \Gamma_F (\alpha'),
\end{align}
and for up to NLL and NLL' accuracy are given by
\begin{align}
K_F(\mu, \mu_0) &=-\frac{\gamma_F^0}{2 \beta_0} \ln r -\frac{2 \pi \Gamma_F^0}{(\beta_0)^2} \Big{\lbrack} \frac{r-1+r\ln r}{\alpha_s(\mu)}
+ \left( \frac{\Gamma^1_c}{\Gamma^0_c}-\frac{\beta_1}{\beta_0} \right) \frac{1-r+\ln r}{4 \pi}+\frac{\beta_1}{8 \pi \beta_0}
\ln^2 r  \Big{\rbrack}, \\
\omega_F(\mu, \mu_0) &= - \frac{\Gamma_F^0}{j_F \beta_0} \Big{\lbrack} \ln r + \left( \frac{\Gamma^1_c}{\Gamma^0_c} -
\frac{\beta_1}{\beta_0}  \right) \frac{\alpha_s (\mu_0)}{4 \pi}(r-1)\Big{\rbrack},
\end{align}
where $r=\alpha(\mu)/\alpha(\mu_0)$ and $\beta_n$ are the coefficients of the QCD $\beta$-function,
\begin{equation}
\beta(\alpha_s) = \mu \frac{d \alpha_s}{d \mu}= -2 \alpha_s \sum_{n=0}^{\infty} \left( \frac{\alpha_s}{4 \pi} \right)^{1+n} \beta_n \, .
\end{equation}

\bibliography{article}

\end{document}